\documentclass[sigplan,nonacm]{acmart}
\usepackage[]{hyperref}

\usepackage{amsmath,amssymb,amsfonts}
\usepackage{algorithm}
\usepackage{algpseudocode}
\usepackage{amsmath}
\algrenewcommand\alglinenumber[1]{\footnotesize #1}

\usepackage{subcaption}  
\usepackage{graphicx}
\usepackage{textcomp}
\usepackage{xcolor}
\usepackage{xspace}
\usepackage{mdframed}
\usepackage{tabularx}
\usepackage{tikz}
\usepackage{enumitem}
\usepackage{array}
\usepackage{color}
\usepackage{soul}
\usepackage{subcaption}
\usepackage{makecell}
\usepackage{multirow}
\usepackage{caption} 
\usepackage{subcaption}   
\usepackage{placeins}     

\usepackage{enumitem}
\setlist{nolistsep} %
\usepackage{pifont} 

\usepackage{xcolor}

\newcommand{\cmark}{\textcolor{green}{\ding{51}}} 
\newcommand{\xmark}{\textcolor{red}{\ding{55}}}           

\newcommand{\ignore}[1]{}

\newcommand{\name}{\text{Nexus}}

\begin{document}
\raggedbottom


\title{\text{\name}: Proactive Intra-GPU Disaggregation of Prefill and
Decode in LLM Serving}



\author{Xiaoxiang Shi}
\authornote{Both authors contributed equally to this work.}
\email{lambda7xx@gmail.com}
\affiliation{%
  \institution{Independent Researcher}
     \country{}
}

\author{Colin Cai}
\authornotemark[1]
\email{cai9@berkeley.edu}
\affiliation{%
  \institution{Independent Researcher}
     \country{}
}

\author{Junjia Du}
\email{junjia001@e.ntu.edu.sg}
\affiliation{%
  \institution{Nanyang Technological University}
     \country{}
}




\author{Zhihao Jia}
\email{zhihao@cmu.edu}

\affiliation{%
  \institution{Carnegie Mellon University}
     \country{}
}

\begin{abstract}

Current prefill–decode (PD) disaggregation is typically deployed at the level of entire serving engines\footnote{We use the term \textit{serving engine} to denote a unit of GPUs that manages exactly one complete copy of the model weights.}, assigning separate GPUs to handle prefill and decode phases. While effective at reducing latency, this approach demands more hardware. To improve GPU utilization, Chunked Prefill mixes prefill and decode requests within the same batch, but introduces phase interference between prefill and decode.

While existing PD disaggregation solutions separate the phases across GPUs, we ask: can the same decoupling be achieved within a single serving engine? The key challenge lies in managing the conflicting resource requirements of prefill and decode when they share the same hardware. In this paper, we first show that chunked-prefill requests cause interference with decode requests due to their distinct requirements for GPU resource. Second, we find that GPU resource exhibits diminishing returns—beyond a saturation point, increasing GPU allocation yields negligible latency improvements. This insight enables us to split a single GPU’s resources and dynamically allocate them to prefill and decode on the fly, effectively disaggregating the two phases within the same GPU.



Across a range of models and workloads, our system \text{\name} achieves up to 2.2$\times$ higher throughput, 20$\times$ lower TTFT, and 2.5$\times$ lower TBT than vLLM, and outperforms SGLang with up to 2$\times$ higher throughput, 2$\times$ lower TTFT, and 1.7$\times$ lower TBT, and achieves 1.4× higher throughput than
vLLM-disaggregation with only half the number of
GPUs.

\end{abstract}

\maketitle

\pagestyle{plain} 

\section{Introduction}
Transformer-based~\cite{Transformers} Large Language Models (LLMs) \cite{ChatGPT,DeepSeek,FewShotLearnging,gemini} have achieved state-of-the-art performance on a wide range of tasks, from natural language understanding to code synthesis~\cite{gp4-report, qwen1,qwen2.5, qwen3, kimi, deepseek_r1,deepseek_v3, deepseek_coder,deepcoder}. The success has also driven their integration into latency-sensitive applications such as chatbots \cite{ChatGPT,gemini,DeepSeek, kimi, gp4-report, deepseek_r1}, search assistants \cite{Perplexity}, and AI-augmented IDEs \cite{Cursor}. In interactive settings, even small delays matter a lot: humans perceive latencies above one second as disruptive \cite{DistServe}, and sub-second improvements has been shown to substantially boost engagement~\cite{GIGASPACE}. Therefore, latency has become a critical performance metric for LLM serving.



LLM inference consists of two distinct stages with heterogeneous resource demands: \textit{prefill} and \textit{decode}. In the prefill stage, the model processes the entire prompt in a single forward pass to produce the first output token while populating the key-value (KV) cache. This stage is typically \textit{compute-bound}~\cite{NanoFlow, Splitwise}, dominated by large matrix multiplications that fill GPU compute units. In contrast, the decode stage generates tokens one at a time, attending to all previously cached KV states. 
The decode stage, processing one token per request in
each forward pass, features lightweight compute but requires
reading the full model weights and the entire KV cache, making
it heavily constrained by memory bandwidth~\cite{NanoFlow, Splitwise, PoD, vLLM-SOSP23}.


These two stages naturally give rise to two critical latency metrics in LLM serving: \textit{time-to-first-token (TTFT)}, the delay before the first output token is produced, determined by the prefill stage; and \textit{Time-between-Tokens (TBT)}, the latency between subsequent tokens, dictated by each iteration of the decode loop. These metrics directly impact the user experience in latency sensitive applications. Optimizing TTFT and TBT has driven a wave of innovations in LLM serving systems, spanning scheduling, batching, and kernel design~\cite{ChunkPrefill,SGlang,vLLM-SOSP23,FastServe,Llumnix,ye2025flashinfer,DistServe,FasterTransformer, specinfer, spotserve,exegpt,helix,PoD,neustream,pensieve,windserve,specee}.

\begin{figure}[t]
  \centering

  \begin{subfigure}[t]{0.48\linewidth}
    \includegraphics[width=\linewidth]{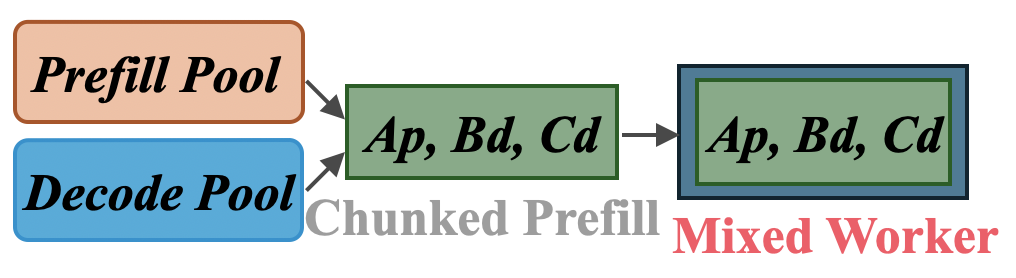}
    \caption{Monolithic.}
    \label{fig:intra_engine_chunk_prefill}
  \end{subfigure}
  \hfill
  \begin{subfigure}[t]{0.48\linewidth}
    \includegraphics[width=\linewidth]{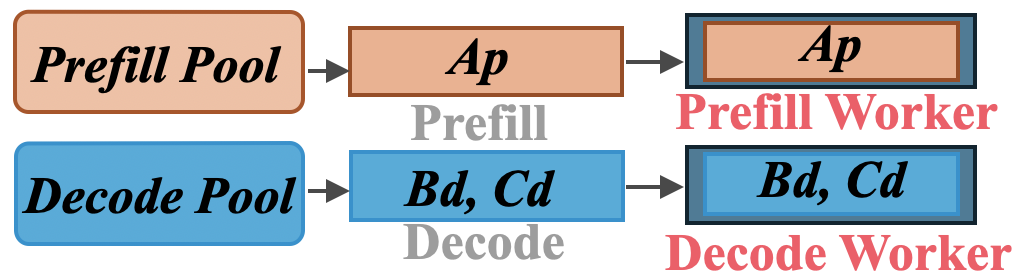}
    \caption{PD Disaggregated.}
    \label{fig:inter-engine_disagg}
  \end{subfigure}

  \vspace{1mm}

    \begin{subfigure}[t]{0.48\linewidth}
      \vspace*{-6mm} 
      \includegraphics[width=\linewidth]{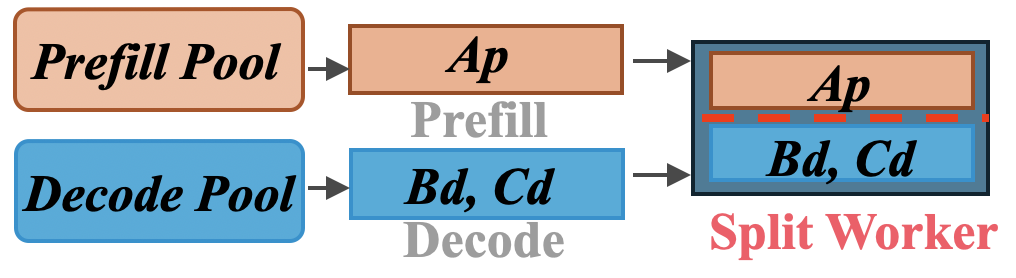}
      \caption{Intra-engine disaggregation.}
      \label{fig:intra_engine_disagg}
    \end{subfigure}
    \hfill
    \begin{subfigure}[t]{0.48\linewidth}
      \centering
      \scriptsize
      \renewcommand{\arraystretch}{1.1}
      \setlength{\tabcolsep}{3pt}
      \begin{tabular}{|c|c|c|c|}
        \hline
        \textbf{System} & \textbf{TTFT} & \textbf{TBT} & \textbf{Util.} \\
        \hline
        Monolithic & \xmark & \xmark & \cmark \\
        Disagg.    & \cmark & \cmark & \xmark \\
        \name{}    & \cmark & \cmark & \cmark \\
        \hline
      \end{tabular}
      \caption{System characteristics.}
      \label{tab:system_characteristics}
    \end{subfigure}
    
  \vspace{1mm}
  \caption{\small \textbf{Design evolution of LLM inference systems.} Comparison between monolithic, disaggregated, and intra-engine disaggregated designs. Ap is the prefill phase of request A; Bd, Cd, and Dd are the decode phases of requests B, C, and D.}
  \vspace{-4mm}
  \label{fig:intro_architecture}
\end{figure}

Existing LLM serving systems fall into two broad classes based on how they place prefill and decode execution (Figure~\ref{fig:intra_engine_chunk_prefill}, \ref{fig:inter-engine_disagg}).

\textit{Monolithic systems}~\cite{vLLM-SOSP23, SGlang, NanoFlow, ChunkPrefill, pensieve,helix, SGLang-v0.4} execute both stages within a single engine. To improve utilization, recent designs like Sarathi-Serve~\cite{ChunkPrefill} adopt \textit{chunked prefill}, where long prompts are split into shorter chunks and batched alongside decode tokens. This improves token throughput and reduces \textsc{TBT} by increasing batch efficiency. However, this design mixes compute heavy prefill and memory sensitive decode operations in the same batch, causing interference between the prefill and decode and increasing TBT (see Section~\ref{subsec:chunk_prefill_interference} for details). 


\textit{Disaggregated systems}~\cite{DistServe, Splitwise, MoonCake,TetriInfer} assign prefill and decode to separate engines, transferring KV cache data between them. This eliminates interference and achieves consistently low \textsc{TTFT} and \textsc{TBT}. But it sacrifices efficiency~\cite{windserve}: decode engines are often underutilized, and multi-GPU deployment incurs additional hardware and communication overhead, especially when KV state is large.

This paper asks a simple question: \textit{Can a single engine achieve low TTFT and TBT without sacrificing GPU utilization?} We answer “yes” by introducing \textit{intra-engine logical disaggregation}, which separates prefill and decode execution within a single serving engine. This design is grounded in three observations. First, mixing compute heavy prefill and memor bound decode in chunked batches creates fine-grained interference that increases \textsc{TBT}. Second, both stages exhibit diminishing returns beyond moderate compute allocations overprovisioning is wasteful, and equal partitioning is inefficient. Third, modern accelerators~\cite{H100, B200} now provide sufficient on device memory and compute to support disaggregation within a single engine, avoiding the communication overheads of cross-GPU designs.

However, enabling this form of disaggregation raises new challenges. Unlike traditional multi-engine systems, we must partition and schedule GPU resources \textit{dynamically within a single serving engine}, while minimizing contention and adapting to workload shifts in \textit{sub-second timescales}. Prefill and decode pressure evolve continuously with prompt length, KV cache footprint, and request mix, requiring fine-grained orchestration that scales.

To address these challenges, we present \textit{\name} (Figure~\ref{fig:intro_architecture}c), a monolithic LLM serving engine that achieves intra-engine prefill–decode disaggregation. \text{\name} is built on three components:  
\textbf{(1)} A lightweight analytical cost model predicts latency as a function of resource allocation, prompt length, and cache usage.  
\textbf{(2)} A greedy search algorithm consults this model to select low-latency resource partitions in real time, requiring only a few closed-form evaluations per update.  
\textbf{(3)} Two phase-specific schedulers, shortest-prompt-first for prefill and FCFS for decode, exploit their differing characteristics to optimize \textsc{TTFT} and \textsc{TBT}.  

Together, these mechanisms allow \text{\name} to match the high utilization of monolithic designs while achieving the isolation benefits of disaggregated systems—without incurring cross-device transfers or relying on additional hardware.

We implement \text{\name} by extending vLLM~\cite{vLLM-SOSP23}, adding fine-grained resource partitioning and concurrent prefill–decode execution within a single engine. \text{\name} runs on commodity GPUs without kernel modification, supports decoder-only LLMs, and requires no specialized hardware.

\vspace{0.5em}
\noindent \textbf{This paper makes the following contributions:}
\begin{itemize}[topsep=0pt, partopsep=0pt, itemsep=0.1em]
    \item  We identify the limitations of existing solutions in serving LLMs and propose intra-engine PD disaggregation as a solution.
    
    \item We develop a lightweight, adaptive scheduling mechanism that enables \textit{sub-second resource repartitioning} and phase-aware prioritization, making intra-engine PD disaggregation practical under real-world, dynamic workloads.
    
    \item We implement \name{} as a drop-in vLLM extension and evaluate it on production-scale LLMs and traffic. \name{} achieves up to \textbf{2.2$\times$} and \textbf{2$\times$} higher throughput, \textbf{20$\times$} and \textbf{2$\times$} lower TTFT, and \textbf{2.5$\times$} and \textbf{1.7$\times$} lower TBT than vLLM and SGLang respectively and outperforms disaggregated vLLM with half the GPU resources.
\end{itemize}

\section{Background}



\begin{figure}[t]
\centering
\includegraphics[width=\linewidth]{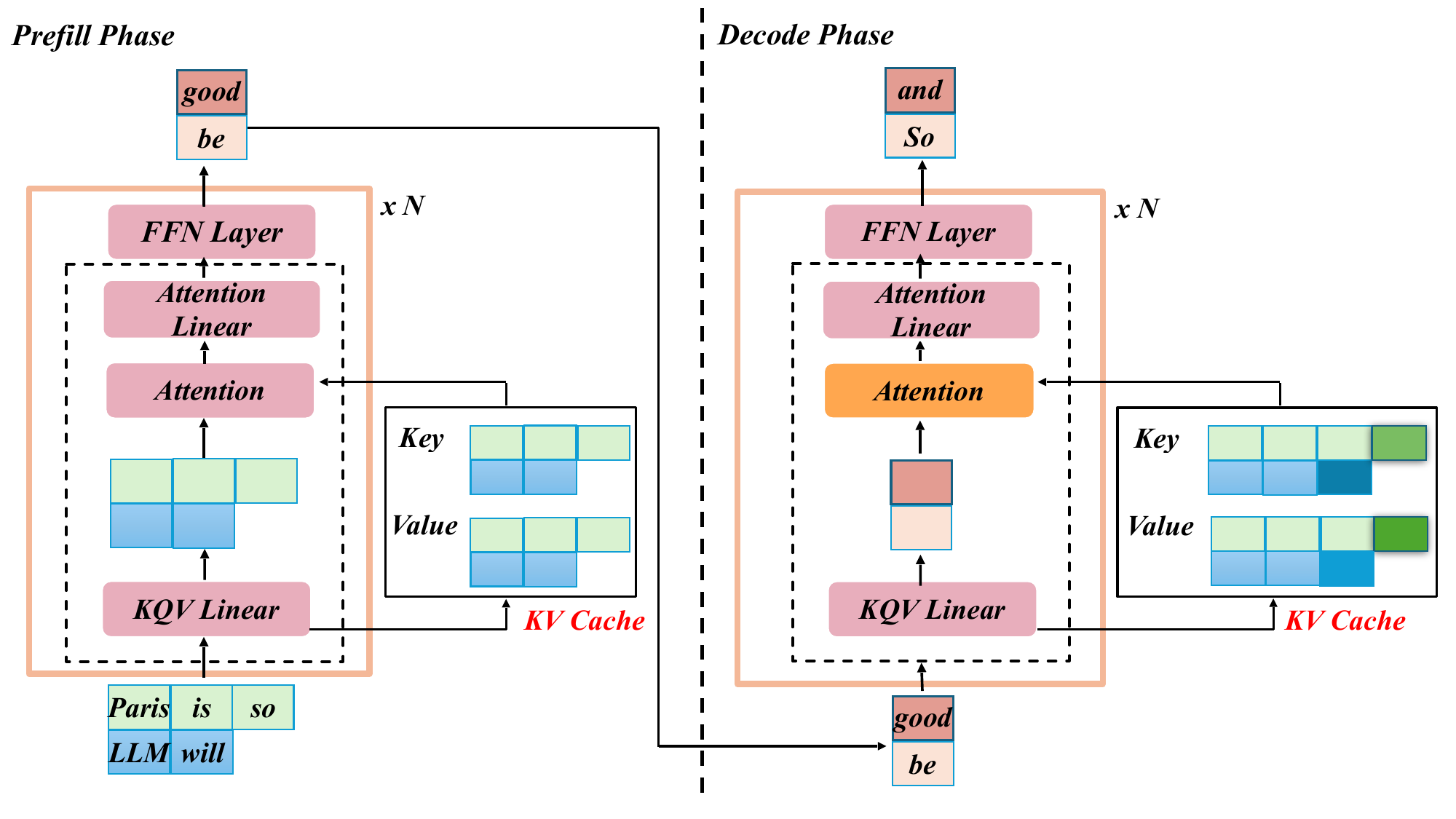}
\vspace{-2mm}
\caption{\small 
\textbf{Inference process of transformer-based LLMs.} 
Red boxes indicate compute-bound operations (KQV Linear, Prefill Attention, Attention Linear, and FFN Layer), while the orange box (Attention) represents a memory-bound operation. Auxiliary components such as LayerNorm are omitted for clarity.
}
\vspace{-4mm}
\label{fig:transformer_layers}
\end{figure}
\subsection{Architecture of Transformer-based LLM}

Most large language models (LLMs)~\cite{Llama3, Qwen, GPT-3} adopt a decoder-only Transformer architecture~\cite{Transformers}, consisting of a stack of layers that apply self-attention and feed-forward networks (FFNs). During inference, LLMs operate in two distinct phases: \textit{prefill} and \textit{decode}. Figure~\ref{fig:transformer_layers} illustrates the computation involved in each phase. The left panel depicts the prefill phase, where the prompt is processed to produce the first token and KV cache. The right panel shows the decode phase, where tokens are generated autoregressively using the cached KV states. Each Transformer layer includes both attention-related operations and dense operations~\cite{NanoFlow}. The latter consist of the linear projections for queries, keys, and values (Q/K/V), the output projection following attention, and the FFN sublayer. All these dense operations are compute bound~\cite{NanoFlow, Splitwise}.

\subsection{Dense Operations} 

\vspace{1mm}


\noindent \textbf{Q/K/V Projections.} Before attention, the input \( X \) is projected by weight matrices \( W_Q \), \( W_K \), and \( W_V \) to produce the corresponding Query, Key, and Value tensors. \( W_Q, W_K, W_V \in \mathbb{R}^{d \times d} \) where d is the hidden size.


\begin{equation}
Q = X W_Q, \quad K = X W_K, \quad V = X W_V
\end{equation}

Without caching, all \( L \) tokens are processed, leading to \( O(L d^2) \) compute. With caching, only the \( n \) new tokens require projection, reducing the cost to \( O(n d^2) \). 

\vspace{1mm}
\noindent \textbf{Attention Output Projection.}
The attention output \( A \in \mathbb{R}^{n \times d} \) is projected using:

\begin{equation}
O = A W_O, \quad W_O \in \mathbb{R}^{d \times d}
\end{equation}

The cost is \( O(n d^2) \), or \( O(L d^2) \) if no caching is used.

\vspace{1mm}
\noindent \textbf{Feed-Forward Network (FFN).}
Each token is independently processed by a two-layer MLP with a non-linear activation (typically GELU):

\begin{equation}
\text{FFN}(x) = \text{GELU}(x W_1 + b_1) W_2 + b_2
\end{equation}

where \( W_1 \in \mathbb{R}^{d \times d_{\text{ff}}} \), \( W_2 \in \mathbb{R}^{d_{\text{ff}} \times d} \), and typically \( d_{\text{ff}} = 4d \). The cost is \( O(n d \cdot d_{\text{ff}}) \), or \( O(L d \cdot d_{\text{ff}}) \) in the absence of caching. Given the size of \( d_{\text{ff}} \), this is usually the most FLOP-intensive component of the layer.





\subsection{Attention Operation}

Self-attention computes token-level dependencies by comparing queries with keys and applying attention weights to values. Given query \( Q \in \mathbb{R}^{n \times d} \), key \( K \in \mathbb{R}^{L \times d} \), and value \( V \in \mathbb{R}^{L \times d} \), attention is computed as:

\begin{equation}
S = \frac{Q K^\top}{\sqrt{d}}, \quad A = \text{softmax}(S) V, \quad A \in \mathbb{R}^{n \times d}
\end{equation}

The attention computation has two main components:
- Computing similarity scores \( S \in \mathbb{R}^{n \times L} \): cost \( O(n L d) \),
- Applying softmax and aggregating over values: cost \( O(n L d) \).

Thus, the overall attention complexity is \( O(n L d) \). In the absence of KV caching, this becomes \( O(L^2 d) \).

\vspace{1mm}
\noindent \textbf{Prefill vs. Decode Attention.}
Assuming cache is enabled, the difference between prefill and decode arises from the size of \( n \). In the \textit{prefill phase}, a chunk of \( n \) new tokens is processed in parallel. The attention computation involves matrix-matrix multiplications and has cost \( O(n L d) \), which is compute-bound~\cite{NanoFlow} and benefits from parallelism. In the \textit{decode phase}, only one new token is processed at a time (\( n = 1 \)), resulting in a matrix-vector multiplication (GEMV) with cost \( O(L d) \). While the FLOP count is low, this operation is memory-bound~\cite{NanoFlow} due to repeated access to the model weights and the growing KV cache.

These contrasting patterns lead to different system bottlenecks~\cite{vLLM-SOSP23, PoD}: prefill benefits from parallelism, while decode demands cache-efficient execution.

\subsection{Batching in LLM Serving}
Modern LLM serving systems must coordinate prefill and decode phases with fundamentally different latency objectives.
Prefill performance directly determines TTFT, while decode responsiveness governs TBT. Since both phases contend for GPU resources. Existing LLM serving systems fall into two broad categories: Monolithic Systems and PD Disaggregation systems, depending on how they manage prefill and decode execution.


\paragraph{Monolithic System} Monolithic system~\cite{vLLM-SOSP23, SGlang} such as Sarathi-Serve~\cite{ChunkPrefill}(Figure~\ref{fig:intra_engine_chunk_prefill}) splits long prompts into fixed-size chunks and mixing them with decode tokens in a shared batch queue, these systems construct mixed-phase batches that improve GPU utilization and reduce head-of-line blocking. 



\paragraph{PD Disaggregation}
To isolate prefill and decode execution, some systems adopt prefill–decode disaggregation~\cite{DistServe, Splitwise, TetriInfer, MoonCake}(Figure~\ref{fig:inter-engine_disagg}), assigning each phase to separate serving engine. This strategy enables independent scheduling and eliminates interference between prefill and decode, particularly in multi-GPU environments.



\begin{figure}[t]
\centering
\includegraphics[width=\linewidth]{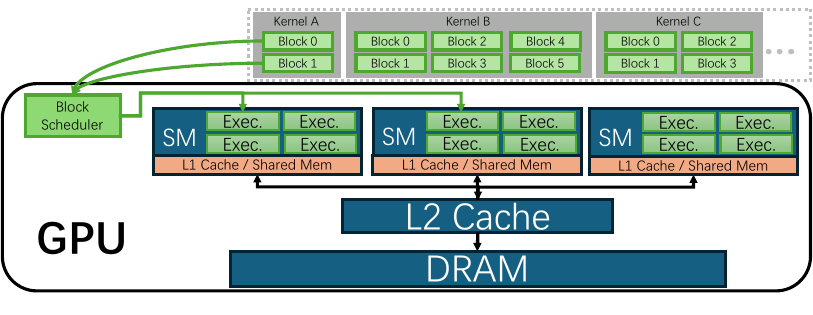}
\vspace{-2mm}
\caption{\small \textbf{Simplified GPU execution model.} 
Modern GPUs share a global kernel queue, with SMs (streaming multiprocessors) dynamically fetching kernels to execute. Concurrently executing kernels compete for shared memory bandwidth.} 
\vspace{-4mm}
\label{fig:gpu_execution_model}
\end{figure}
\subsection{GPU Execution Model}

Modern LLM inference relies heavily on GPU acceleration. Figure~\ref{fig:gpu_execution_model} presents a simplified view of GPU execution, abstracting low-level scheduling mechanisms in favor of architectural components most relevant to LLM serving workloads.

GPUs consist of multiple Streaming Multiprocessors (SMs), each with its own block executor, L1 cache, and register file, while sharing a unified L2 cache and off-chip DRAM. Kernels are submitted via a global software queue and consists of many blocks. The blocks carry operations and are scheduled onto available executors in SMs by a hardware-level scheduler, which is largely opaque to software and does not provide preemptive control or fine-grained prioritization.

\section{Motivation}
Modern LLM serving must balance two asymmetric phases: compute-heavy prefill and memory-bound decode. In the following section, we examine limitations of current systems that mix or isolate these phases (\S\ref{subsec:chunk_prefill_interference}), explore how both exhibit diminishing returns with added resources (\S\ref{subsec:diminishing_return}), and highlight the inefficiency of static resource partitioning under memory bandwidth contention (\S\ref{subsec:contention}). Together, these insights motivate intra-engine disaggregation with  GPU resourcereallocation on the fly.

\smallskip
\noindent\textbf{Setup.} Unless stated otherwise, we evaluate with Long Data Collections Workload (\S\ref{subsec:experimental_setup}) using Qwen2.5–3B on a single NVIDIA L20 GPU. Requests follow a Poisson arrival at 2.5 req/s. We use NVIDIA MPS~\cite{MPS} to control SM partitioning.

\begin{figure}[!t]
\centering
\small
\begin{subfigure}[t]{\columnwidth}
    \centering
    \setlength{\tabcolsep}{3pt}
    \renewcommand{\arraystretch}{1.1}    
    \begin{tabular}{@{}lcccc@{}}
    \toprule
    \textbf{Type} & \textbf{Avg Time(s)}  & \textbf{Count} & \textbf{\%} \\
    \midrule
    Prefill-only  & 0.132  & 2    & 0.02\% \\
    Decode-only   & 0.015    & 563  & 5.80\% \\
    Mixed         & 0.251  & 9150 & 94.18\% \\
    \bottomrule
    \end{tabular}
    \caption{Statistics by Batch Types}
    \label{tab:chunked_prefill_profile}
\end{subfigure}
\hfill
\begin{subfigure}[t]{\columnwidth}
    \centering
    \includegraphics[width=\linewidth]{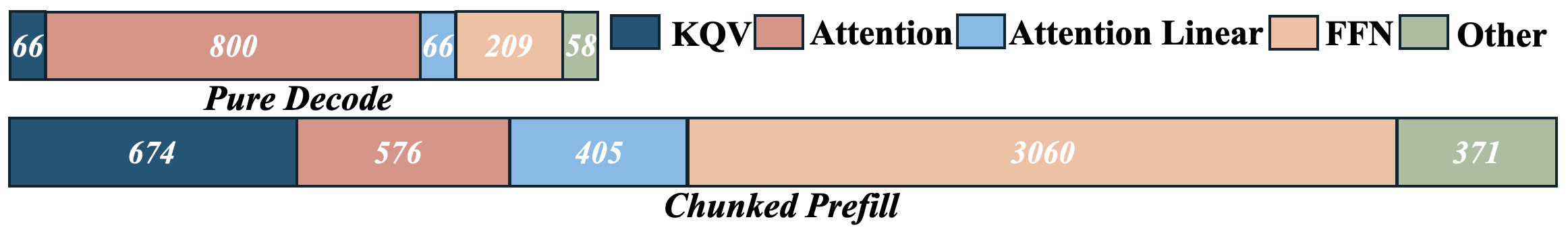}
    \caption{Latency breakdown by kernel.}
    \label{fig:chunk_prefill_decode_kernel_latency}
\end{subfigure}
\vspace{1mm}
\caption{\small \textbf{Latency impact of mixed prefill–decode batches.} 
(a) Prefill-only and decode-only batches have predictable latency, but mixed batches cause 8×–10× slowdown due to interference. 
(b) Kernel-level profiling reveals that even lightweight decode kernels experience inflated runtimes when co-executed with prefill. 
This highlights the fine-grained contention caused by chunked batching.}
\label{fig:chunked_prefill_analysis}
\vspace{-4mm}
\end{figure}
\subsection{Limitation of Existing Solutions}
\label{subsec:chunk_prefill_interference}
Existing LLM serving systems can fall into two architectural categories: \textit{monolithic execution} and \textit{disaggregation}.

\textbf{Disaggregated systems}~\cite{DistServe, MoonCake, Splitwise} places prefill and decode to separate engines, eliminating resource contention and stabilizing latency. However, the clean separation come with steep hardware costs: multiple engines must maintain the full model replica, prefill engine's memory is wasted, and decode often underutilize their assigned GPUs. Worse, coordination is non-trivial. Under dynamic workloads, KV cache eviction and recomputation have been shown to severely inflate both \textsc{TTFT} and \textsc{TBT}~\cite{windserve}.

\textbf{Monolithic systems}~\cite{vLLM-SOSP23, SGlang, ChunkPrefill, pensieve} adopt chunked prefill, batching prefill and decode requests to improve utilization. Although the strategy increases throughput, it does not account for the distinct compute and memory behavior of each phase, causing phase interference. To quantify this, we categorize batches into \textit{prefill-only}, \textit{decode-only}, and \textit{mixed}, and measure their iteration times (Figure~\ref{tab:chunked_prefill_profile}). Despite having similar token counts, mixed batches average 250\,ms, compared to just 15\,ms for decode-only batches.

This slowdown arises from full prefill computation (e.g., KQV projection, FFN) blocking lightweight decode kernels in the same batch. As shown in Figure~\ref{fig:chunk_prefill_decode_kernel_latency}, linear kernel latency in mixed batches is up to 10$\times$ higher than in decode-only batches. Since decode cannot proceed until all prefill kernels are completed, this inflates \textsc{TBT} by more than 8$\times$.

\begin{mdframed}[backgroundcolor=gray!7, linewidth=0.8pt]
\noindent\textbf{\underline{Insight 1.}} 
Chunked prefill improves utilization but introduces fine-grained interference within batches. PD disaggregation avoids this but wastes resources. We aim to achieve both isolation and efficiency within a single engine.
\end{mdframed}

\begin{figure}[t]
  \centering

  \begin{subfigure}[t]{0.32\linewidth}
    \centering
    \includegraphics[width=\linewidth]{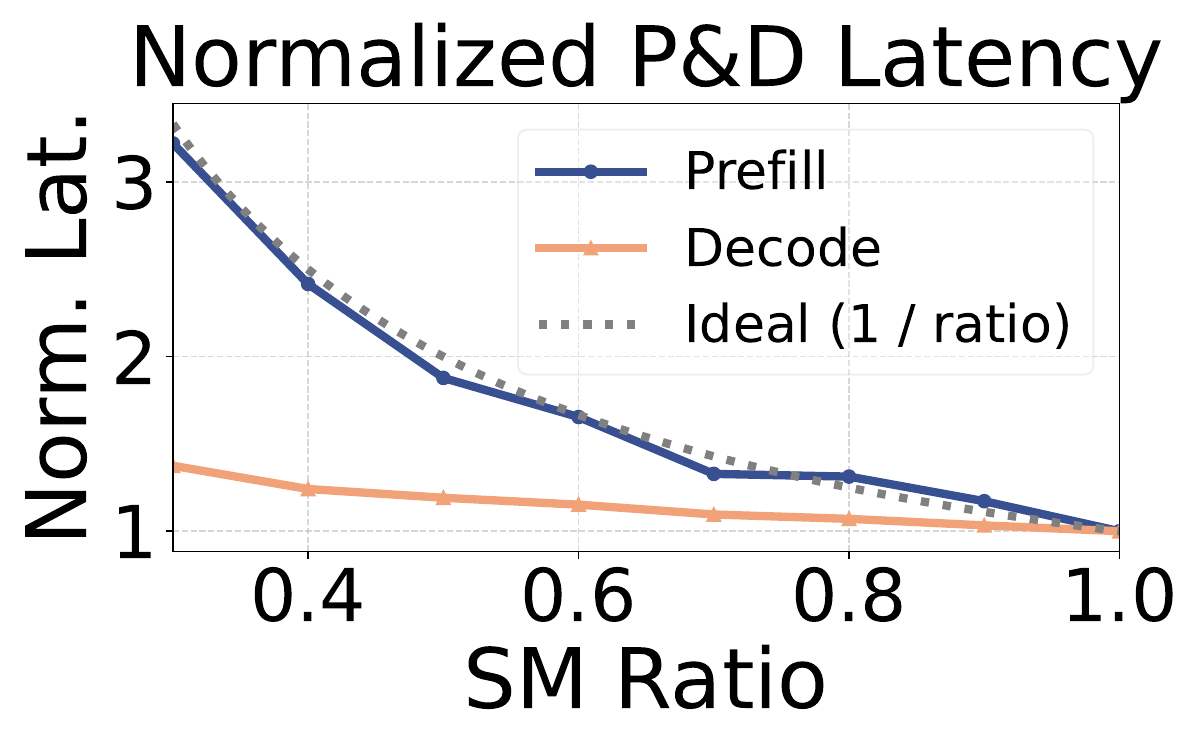}
    \caption{\footnotesize Normalized latency.}
    \label{fig:combined_latency}
  \end{subfigure}
  \hfill
  \begin{subfigure}[t]{0.32\linewidth}
    \centering
    \includegraphics[width=\linewidth]{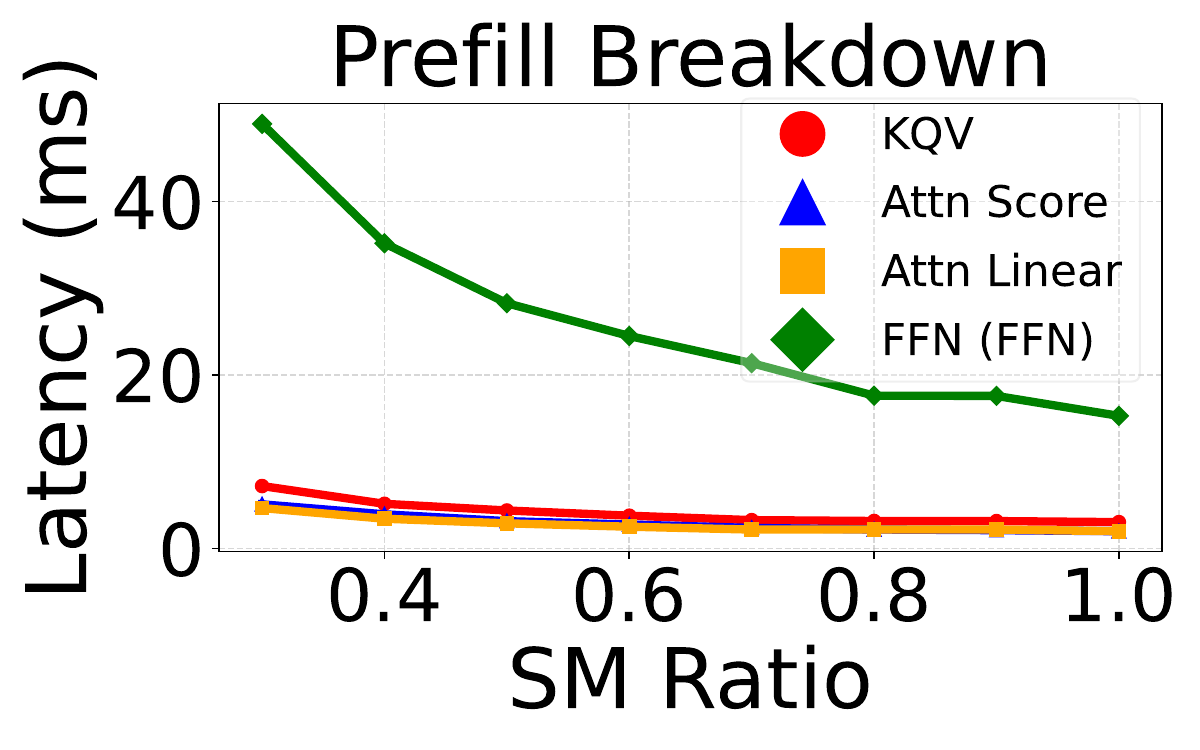}
    \caption{\footnotesize Prefill breakdown.}
    \label{fig:prefill_breakdown}
  \end{subfigure}
  \hfill
  \begin{subfigure}[t]{0.32\linewidth}
    \centering
    \includegraphics[width=\linewidth]{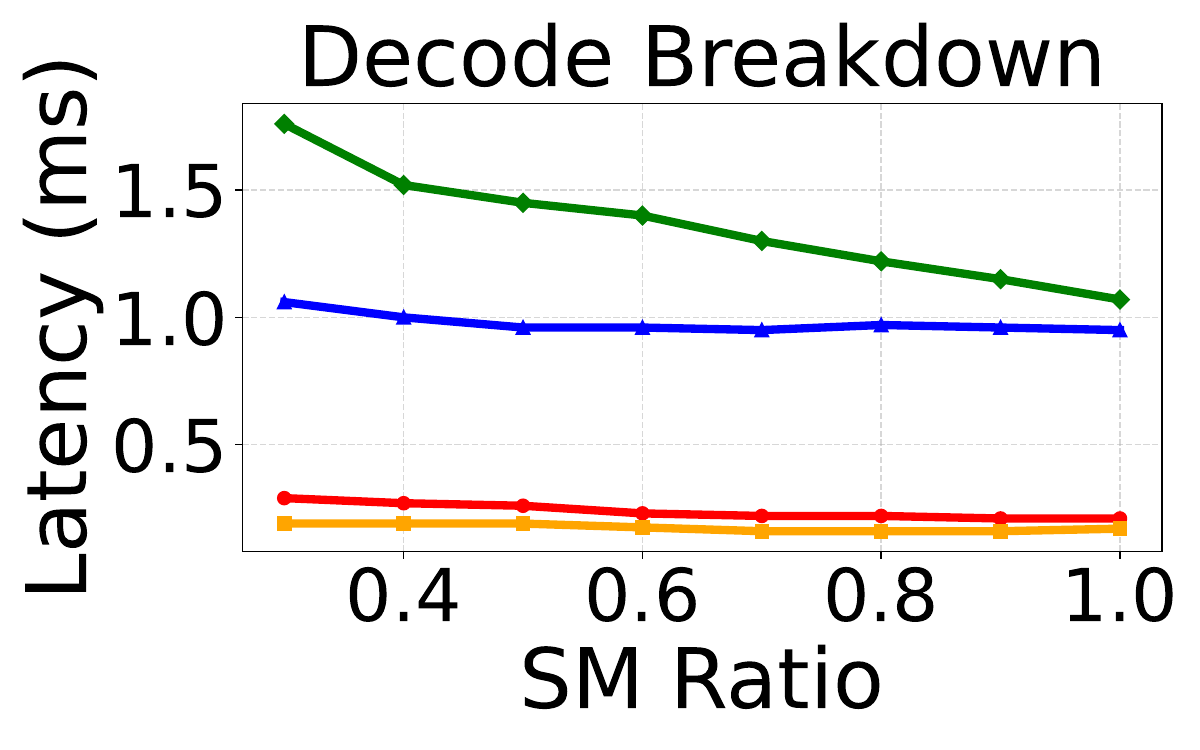}
    \caption{\footnotesize Decode breakdown.}
    \label{fig:decode_breakdown}
  \end{subfigure}

  \caption{\small 
    \textbf{Diminishing returns in prefill and decode with increasing SM allocation.} 
    (a) End-to-end latency for prefill and decode flattens well before full SM usage. 
    (b) Prefill kernels (e.g., FFN, KQV, attention linear) show varied sensitivity to SM scaling, with FFN benefiting the most. 
    (c) Decode kernels saturate quickly, confirming that decode is memory-bound and gains little from additional compute.}
  \label{fig:llm_marginal_breakdown}
  \vspace{-4mm}
\end{figure}

\subsection{Diminishing Returns in Compute Allocation}  
\label{subsec:diminishing_return}
While the Section~\ref{subsec:chunk_prefill_interference} highlights the limitations of both chunked prefill and inter-engine PD disaggregation, we take a step further by exploring an intra-engine PD approach through GPU resource sharing. In this design, prefill and decode are assigned different portions of the same GPU. To allocate GPU resources effectively under prefill–decode separation, we analyze how each phase scales with compute in isolation. We run pure prefill and decode batches under varying SM ratios, measuring both end-to-end latency and per-kernel runtimes.

As shown in Figure~\ref{fig:llm_marginal_breakdown}, prefill latency closely follows the idealized scaling model \( T \propto \frac{1}{r} \), with diminishing returns emerging gradually. For instance, increasing SM allocation from 30\% to 40\% reduces latency by over 25\%, but the gain drops to just 10\% between 70\% and 80\%. Further investigation in Figure~\ref{fig:prefill_breakdown} shows that compute-heavy layers such as FFN continue to benefit, while others such as KQV and projection flatten out earlier, around 60\%.

Decode exhibits much sharper diminishing returns. Increasing SMs from 30\% to 40\% improves latency by only 10\%, and beyond 50\%, additional SMs yield less than 3\% improvement per 10\% increment. This behavior is expected given decode is memory-bound and fails to utilize added SMs effectively. Figure~\ref{fig:decode_breakdown} confirms this: attention and projection layers show minimal runtime reduction with more compute.

These trends suggest a key systems insight: instead of time-slicing all SMs between the two stages, we can spatially partition them to avoid overprovisioning either phase. Notably, disaggregated systems, which allocate entire GPUs to each phase, operate on the far right of these curves, where additional compute offers diminishing marginal benefit.

\vspace{0.3em}
\begin{mdframed}[backgroundcolor=gray!7, linewidth=0.8pt]
\noindent\textbf{\underline{Insight 2.}}  
Both prefill and decode saturate well before full GPU allocation. To improve efficiency, systems should allocate only the SMs needed to meet each phase's demand.
\end{mdframed}

\begin{figure}[!t]      
  \centering
  \begin{subfigure}{0.49\columnwidth}
    \centering
    \includegraphics[width=\linewidth]{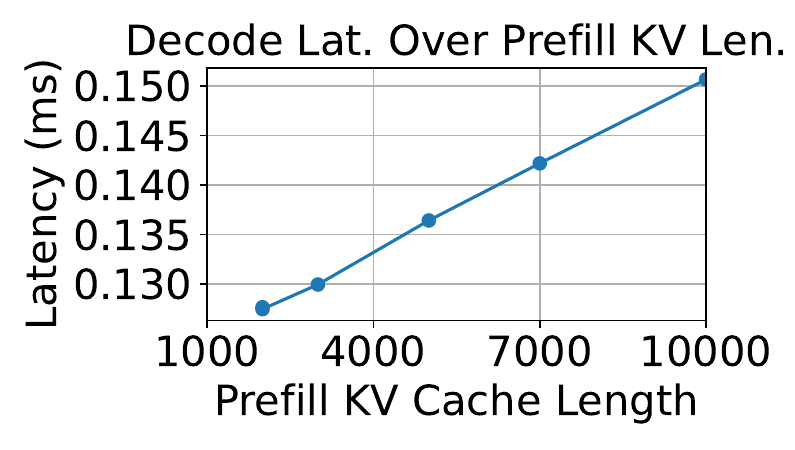}    
    \caption{Impact of prefill KV length on decode latency.}
    \label{fig:attn_contention}
  \end{subfigure}
  \hfill
  \begin{subfigure}{0.49\columnwidth}
    \centering
    \includegraphics[width=\linewidth]{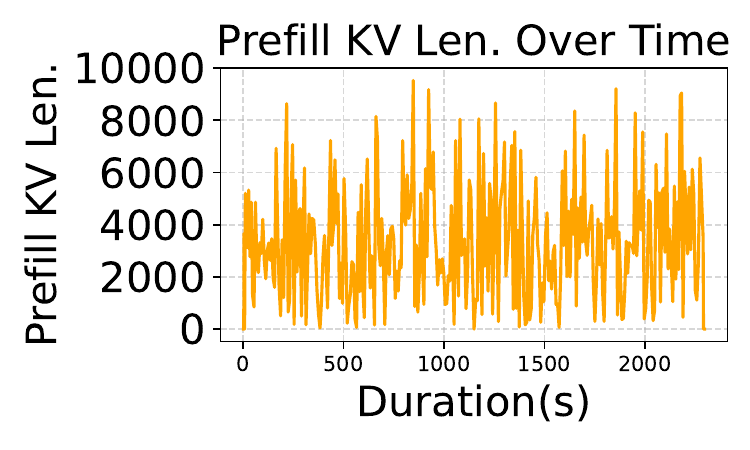}
    \caption{Observed variation in prefill KV length during execution.}
    \label{fig:prefill_kv_read_across_time}
  \end{subfigure}

  \caption{\small \textbf{Memory contention's impact and variability.} (a) Decode latency increases as prefill KV length grows due to shared memory bandwidth pressure; (b) Prefill KV length fluctuates significantly over time, making contention difficult to predict statically.}
\label{fig:memory_contention}
\vspace{-4mm}
\end{figure}

\subsection{Limitation of Static Partitioning}  
\label{subsec:contention}
Section~\ref{subsec:diminishing_return} shows that prefill and decode phases exhibit sharply diminishing returns with increasing SM allocation, motivating finer-grained partitioning. However, even a well-chosen \emph{static} SM partition is suboptimal at runtime. This is because compute demand, while often estimable from inputs like chunk size or sequence length, does not capture dynamic memory behavior. Prefill and decode contend for bandwidth in ways that depend on evolving KV cache sizes, prompt distributions, and decode lengths that emerge only at runtime.

To illustrate this, we co-execute chunks of prefill\footnote{Following Sarathi-Serve~\cite{ChunkPrefill}, long prompts are split into smaller segments and executed alongside decode requests. These chunks of prefill tasks read and write to the KV cache.} and pure decodes under a fixed SM partition. As we can see from Figure~\ref{fig:attn_contention}, increasing the prefill KV length from 2000 to 10000 also increases the latency of the exact same decode batch by 36\%. This slowdown stems from memory bandwidth contention: prefill attention layers perform large KV reads, which overlap with the decode stage's latency-critical memory access.

Moreover, Figure~\ref{fig:prefill_kv_read_across_time} shows that prefill memory traffic is highly irregular, fluctuating significantly over time and making contention both difficult to predict and workload dependent.

These highlight a core limitation: even when compute demands are static and analyzable, bandwidth pressure varies at runtime due to asynchronous prefill/decode evolution. Static SM partitioning fails to respond to this variation, leading to avoidable contention and degraded latency.

\vspace{0.3em}
\begin{mdframed}[backgroundcolor=gray!7, linewidth=0.8pt]
\noindent\textbf{\underline{Insight 3.}}  
Logical PD disaggregation with static partitioning is insufficient. To adapt to emergent memory contention and shifting runtime demands, systems must have fine-grained, dynamic SM reallocation.
\end{mdframed}

\begin{figure}[t]
\centering
\includegraphics[width=\linewidth]{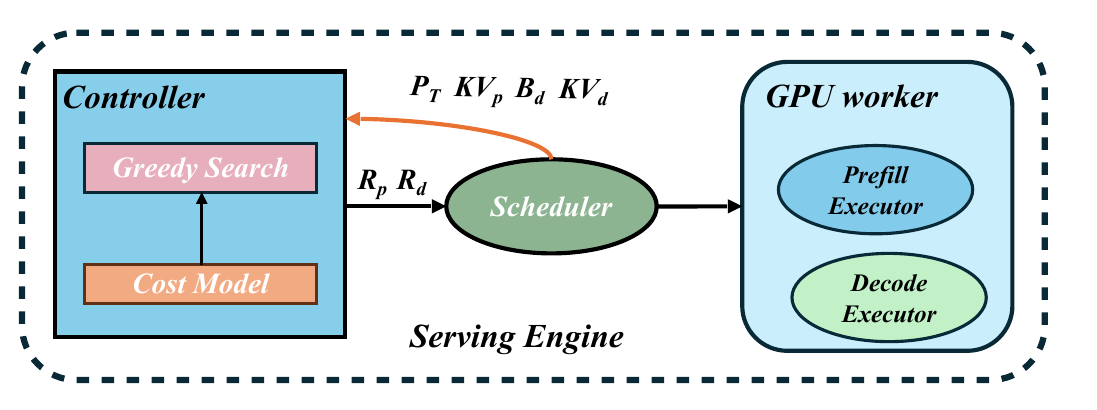}
\vspace{-2mm}
\caption{\small \textbf{System architecture of \name.} $R_p$ and $R_d$ denote the SM ratios allocated to prefill and decode, respectively. $P_T$ represents the chunk prompt length of prefill, $KV_p$ is the total KV cache used during prefill, $B_d$ is the decode batch size, and $KV_d$ is the total KV cache used during decoding.}
\vspace{-4mm}
\label{fig:overview}
\end{figure}

\section{Design}
\label{sec:design}
We present \name{}, a lightweight monolithic LLM serving system that enables \emph{intra-engine prefill–decode disaggregation}. Unlike prior systems that either co-batch prefill and decode or isolate them across GPUs, \name{} partitions GPU resources dynamically, executing both phases concurrently but independently within a single engine.

As shown in Figure~\ref{fig:overview}, \name{} introduces three core mechanisms to enable this fine-grained separation:

\begin{itemize}[topsep=2pt, itemsep=1pt, leftmargin=10pt]
    \item \textbf{Dynamic SM Partitioning} (\S\ref{sec:our_cost_model}): A runtime cost model estimates per-phase iteration latency based on compute scaling and memory contention. We formalize a dual-objective optimization problem and solve it efficiently via a greedy search.
    
    \item \textbf{Stability Control} (\S\ref{sec:repartitioning}): To reduce the overhead of frequent re-partitioning, we apply a hysteresis-style buffer zone that filters out insignificant SM ratio changes.
    
    \item \textbf{Phase-Specific Scheduling} (\S\ref{sec:phase_scheduler}): With prefill and decode isolated, we deploy customized schedulers to optimize TTFT and TBT jointly.
\end{itemize}

These components form a closed-loop control system that continuously adapts GPU resource allocation to match workload demands, preserving high utilization without reintroducing interference.

\subsection{Dynamic SM Partitioning via Cost Model}
\label{sec:our_cost_model}
To execute prefill and decode concurrently without introducing significant interference or wasting resources, \name{} must decide at runtime how to split GPU computes between the two phases. This is challenging: exhaustive search is too slow for inference loops, and simple rules fail under dynamic memory contention.

To solve this, \name{} combines three components: 
(1) an analytical cost model that predicts latency under any SM split, 
(2) a dual-mode optimization objective that shifts focus based on runtime signals, and 
(3) a greedy search algorithm that efficiently selects SM partitions with just a few cost model queries.

\subsubsection{Cost Model.}
\label{subsec:latency-modeling}
\name's cost model estimates the latency of prefill and decode under SM allocations without execution, enabling rapid exploration of latency tradeoffs.

Each iteration consists of multiple operators with different bottlenecks. For example FFNs are compute-bound~\cite{NanoFlow}, while adecode ttention may be memory-bound~\cite{NanoFlow} as KV grows. These characteristics shift with workload, so we model each phase’s latency as a sum over operators:
\begin{align}
    T_{\text{prefill}} &= \sum_{i \in \text{PrefillOps}} \max\left(T_i^{\text{compute}}, T_i^{\text{mem}}\right) \\
    T_{\text{decode}} &= \sum_{j \in \text{DecodeOps}} \max\left(T_j^{\text{compute}}, T_j^{\text{mem}}\right)
\end{align}

This operator-level modeling captures shifting bottlenecks, such as decode attention flipping between compute- and memory-bound, without collapsing structure as coarse stage-level models would.

\vspace{0.5em}
\noindent\textbf{Compute Latency.}  
We estimate the compute latency of each operator $o \in \texttt{PrefillOps} \cup \texttt{DecodeOps}$ based on its FLOP count $c_o$ and the SM ratio $r$ assigned to its stage. While latency ideally scales inversely with compute share ($1/r$), real-world performance can deviates depending on kernel(\S\ref{subsec:diminishing_return}).

To model this, we use a two-regime saturation–decay curve
\begin{itemize}
\item \textit{Sub-saturation:} Latency scales near-inversely with $r$ until a saturation threshold $R_{\text{sat}}$;
\item \textit{Post-saturation:} Additional SMs yield diminishing returns, modeled by a decay coefficient $\lambda$.
\end{itemize}

\begin{equation}
T_o^{\text{compute}}(c_o, r) = 
\begin{cases}
\displaystyle \frac{c_o}{r \cdot C} & \text{if } r \leq R_{\text{sat}} \\
\displaystyle \frac{c_o}{R_{\text{sat}} \cdot C} \cdot \left(1 + \lambda \cdot (r - R_{\text{sat}})\right) & \text{otherwise}
\end{cases}
\end{equation}
where $C$ is the peak throughput of the GPU. 

We extract $R_{\text{sat}}$ and $\lambda$ per operator from end-to-end measurements of the full stage (prefill or decode) under varying SM allocations. 

\vspace{0.5em}
\noindent\textbf{Memory Access Latency.}  
As shown in \S\ref{subsec:contention}, memory contention can significantly affect latency when prefill and decode execute concurrently on shared hardware. To capture this effect, we model decode memory latency as a function of (1) temporal overlap with prefill, and (2) the relative memory bandwidth demands of each phase.

Let $T_{\text{prefill}}$ denote total prefill duration, and $T_{\text{prefill}}^{\text{attn}}$ be the estimated time on memory-bound attention layers. Then the probability that decode overlaps with prefill attention is:
\begin{align}
    P_{\text{attn}} = \frac{T_{\text{prefill}}^{\text{attn}}}{T_{\text{prefill}}}
\end{align}

We conservatively assume that prefill’s dense layers consume memory bandwidth during the remaining time, yielding:
\[
P_{\text{dense}} = 1 - P_{\text{attn}}
\]

Assuming full bandwidth saturation during each overlap window, we allocate effective bandwidth to decode based on its share of memory traffic. Let $m_d$ be the total memory bytes accessed by decode attention, $m_{p1}$ the bytes accessed by prefill attention, and $m_{p2}$ those accessed by prefill's dense operators.\footnote{We estimate memory access volume for linear layers using known parameter sizes from the model architecture. For attention, we calculate KV memory traffic based on the number of cached tokens tracked by vLLM, multiplied by the per-token key/value size determined from model dimensions.} Then the effective bandwidth for decode is:
\[
B_{\text{decode}} =
  \frac{m_d}{m_d + m_{p1}} \cdot P_{\text{attn}} \cdot B +
  \frac{m_d}{m_d + m_{p2}} \cdot (1 - P_{\text{attn}}) \cdot B
\]

Decode memory latency is then computed as:
\begin{align}
    T_{\text{decode}}^{\text{mem}} = \frac{m_d}{B_{\text{decode}}}
\end{align}

This formulation captures two important dynamics:
(1) contention grows with total memory traffic, reducing effective bandwidth; 
(2) allocating more SMs to decode slows prefill (via compute contention), stretching $T_{\text{prefill}}$, which reduces $P_{\text{attn}}$ and mitigates contention. This feedback is integrated into the overall cost model and guides SM partitioning decisions.

While decode involves multiple operators, we model memory contention only for attention, which dominates bandwidth usage. Other components are lightweight or compute-bound, and do not significantly impact contention. For prefill, we estimate memory latency assuming peak bandwidth, and use the resulting memory-bound segments only to compute $P_{\text{attn}}$. This separation avoids circular dependencies while capturing the dominant interaction between phases.

\subsubsection{Optimization Objective.}  
\label{subsec:objective}
Given the cost model, \name{} selects an SM partition that balances performance and memory pressure. However, since prefill and decode run concurrently and compete for resources, optimizing both simultaneously is infeasible.

To resolve this, \name{} formulates a \emph{dual-objective latency optimization}: prioritize one phase while constraining the other to remain within a slowdown budget. This allows flexible tradeoffs between TBT and TTFT, depending on workload state. The choice is also guided by runtime signals, such as KV cache usage, to avoid pathological behaviors like OOM.

\vspace{0.5em}
\noindent\textbf{Formulation.}
Let $T_{\text{prefill}}(R_p)$ and $T_{\text{decode}}(R_d)$ denote the estimated latencies under a given SM split $R_p$ and $R_d = 1 - R_p$. We define two optimization modes:

\begin{itemize}
    \item \textbf{Decode-prioritized:}
    \[
    \begin{aligned}
    \min_{R_p} \quad & T_{\text{decode}}(1 - R_p) \\
    \text{s.t.} \quad & T_{\text{prefill}}(R_p) \leq \alpha \cdot T_{\text{prefill}}^{\text{min}} \\
    & 0 \leq R_p \leq 1
    \end{aligned}
    \]

    \item \textbf{Prefill-prioritized:}
    \[
    \begin{aligned}
    \min_{R_p} \quad & T_{\text{prefill}}(R_p) \\
    \text{s.t.} \quad & T_{\text{decode}}(1 - R_p) \leq \beta \cdot T_{\text{decode}}^{\text{min}} \\
    & 0 \leq R_p \leq 1
    \end{aligned}
    \]
\end{itemize}
Here, $T^{\text{min}}$ denotes the ideal latency when a stage is allocated all SMs, and $\alpha, \beta > 1$ are slack variables controlling tolerable slowdowns in the non-prioritized stage. 

\vspace{0.5em}
\noindent\textbf{Runtime Switching.}
We select the objective mode based on live KV cache usage $KV_u$: when $KV_u$ is low, favoring prefill accelerates prompt ingestion; when $KV_u$ is high, prioritizing decode helps reduce memory pressure by completing generations and evicting KV. This feedback mechanism enables resource allocation to respond to workload phases and memory constraints.

\[
\text{Objective Mode} =
\begin{cases}
\text{Prefill-prioritized} & \text{if } KV_u \leq KV_{\text{switch}} \\
\text{Decode-prioritized} & \text{otherwise}
\end{cases}
\]

This mode-switching behavior is key to handling dynamic workloads. Rather than rely on fixed priorities or static thresholds, \name{} adapts its scheduling objective based on system state, enabling both high throughput and memory safety.

\begin{algorithm}[t]
  \small
  \caption{\text{\name}'s SM Partitioning with Greedy Search and Buffer Control}
  \label{algorithm:greedy_search}
  \begin{algorithmic}[1]
    \State \textbf{Input:} $KV_u,\; R_p^{\text{cur}},\; R_d^{\text{cur}}$
    \State \textbf{Output:} New partition $(R_p^{\text{new}}, R_d^{\text{new}})$

    \Procedure{PartitionController}{$KV_u,\; R_p^{\text{cur}},\; R_d^{\text{cur}}$}
      \If{$KV_u > KV_{\text{switch}}$}
        \State $(R_p^{\text{new}}, R_d^{\text{new}}) \gets$ \Call{AdjustPartition}{\textbf{decode}, $R_p^{\text{cur}}, R_d^{\text{cur}}$}
      \Else
        \State $(R_p^{\text{new}}, R_d^{\text{new}}) \gets$ \Call{AdjustPartition}{\textbf{prefill}, $R_p^{\text{cur}}, R_d^{\text{cur}}$}
      \EndIf

      \Statex \Comment{Buffer zone check to suppress unstable or small changes}
      \If{$|R_p^{\text{new}} - R_p^{\text{cur}}| < \delta$}
        \State \Return $(R_p^{\text{cur}}, R_d^{\text{cur}})$
      \Else
        \State \Return $(R_p^{\text{new}}, R_d^{\text{new}})$
      \EndIf
    \EndProcedure

    \Statex
    \Procedure{AdjustPartition}{\textit{target}, $R_p^{\text{cur}}, R_d^{\text{cur}}$}
      \State Let \textit{other} $\gets$ (target is prefill? decode : prefill)
      \State Let Slack $\gets$ (target is prefill? $\beta$ : $\alpha$)
      \State $T_{\text{other}}^{\text{opt}} \gets \textsc{CostModel}(\textit{other}, 100)$
      \State $R_{\text{cur}} \gets$ (target is prefill? $R_p^{\text{cur}}$ : $R_d^{\text{cur}}$)
      \State $R \gets R_{\text{cur}}$
    
      \Statex \Comment{Phase 1: Decrease until constraint is satisfied}
      \While{$\textsc{CostModel}(\textit{other}, 100 - R) > \text{Slack} \cdot T_{\text{other}}^{\text{opt}}$}
        \State $R \gets R - 1$
      \EndWhile
    
      \Statex \Comment{Phase 2: Increase target share until constraint is at limit}
      \While{$R < 100$}
        \State $T_{\text{other}} \gets \textsc{CostModel}(\textit{other}, 100 - (R + 1))$
        \If{$T_{\text{other}} > \text{Slack} \cdot T_{\text{other}}^{\text{opt}}$}
          \State \textbf{break}
        \EndIf
        \State $R \gets R + 1$
      \EndWhile
    
      \State \Return (target is prefill? $(R, 100 - R)$ : $(100 - R, R)$)
    \EndProcedure

  \end{algorithmic}
  \footnotesize\textit{Note:} 
  $\textsc{CostModel}(\text{phase}, R)$ estimates phase latency under $R$;
  all ratios are expressed as percentages of total SMs;
  $\beta$ is tolerance for primary objective's deviation from optimal;
  $\delta$ is the buffer to avoid frequent switches.
\end{algorithm}

\subsubsection{Greedy SM Search}
\label{subsec:greedy-search}
Since prefill and decode iterations run in sub-second times, we do not attempt to globally solve the constrained optimization; instead, we use a two-phase greedy adjustment that is fast, robust, and effective in practice.

\vspace{0.5em}
\noindent\textbf{Phase 1: Constraint Satisfaction.}  
Starting from the current allocation, the algorithm reduces the SM share of the prioritized stage until the non-prioritized stage’s latency constraint is satisfied  (lines 21–23 in Algorithm~\ref{algorithm:greedy_search}).

\vspace{0.5em}
\noindent\textbf{Phase 2: Target Optimization.}  
Once within the feasible region, the algorithm gradually increases the SM share of the prioritized stage, improving its latency as long as the constraint remains satisfied (lines 24–30).

\vspace{0.5em}
\noindent\textbf{Efficiency.}  
The search typically converges within 2–4 cost model evaluations and imposes negligible latency overhead. This fast feedback loop allows \name{} to adapt to runtime contention and shifting workloads.

\begin{figure}[t]
  \centering
  \begin{subfigure}[t]{\linewidth}
    \includegraphics[width=\linewidth]{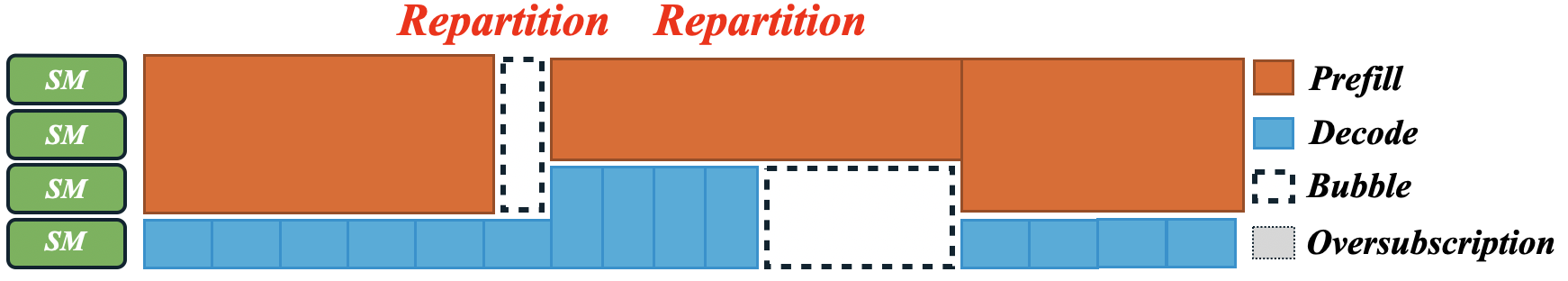}
    \caption{Synchronous solution.}
    \label{fig:switch_sync_design}
  \end{subfigure}
  \vspace{1mm}
  \begin{subfigure}[t]{\linewidth}
    \includegraphics[width=\linewidth]{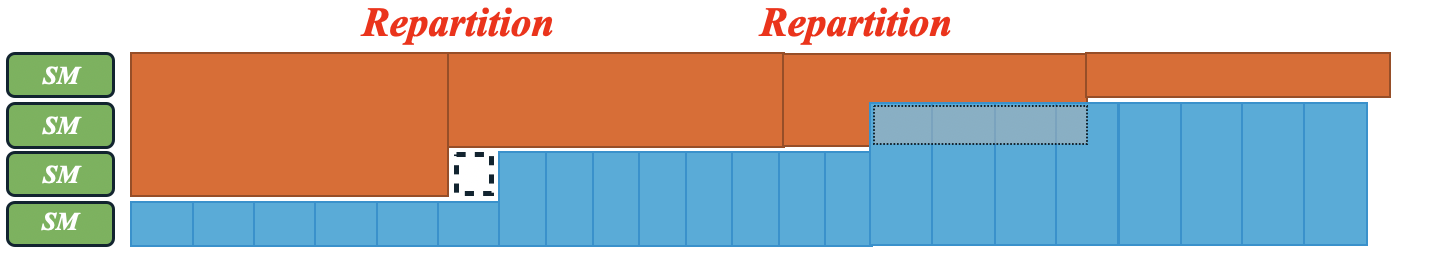}
    \caption{Naive asynchronous solution.}
    \label{fig:switch_async_design}
  \end{subfigure}
  \vspace{1mm}
  \begin{subfigure}[t]{\linewidth}
    \includegraphics[width=\linewidth]{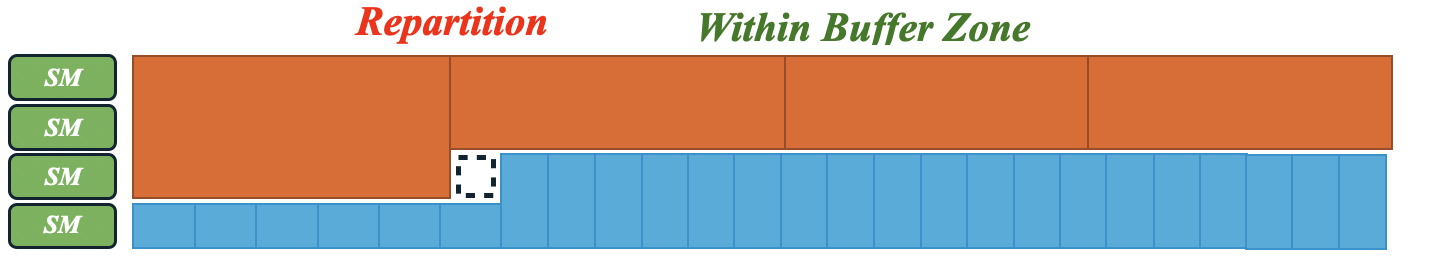}
    \caption{\text{\name} combines asynchronous switching and hysteresis.}
    \label{fig:switch_async_buffered_design}
  \end{subfigure}
  \begin{subfigure}[t]{\linewidth}
\centering
 \end{subfigure}
  \vspace{-2mm}
  \caption{\small \textbf{Mechanisms for SM partition switching.} Comparison between synchronous, asynchronous, and our asynchronous with hysteresis approach.}
  \vspace{-4mm}
  \label{fig:buffered_switch_design}
\end{figure}
\subsection{Hysteresis-Based SM Repartitioning for Stability}
\label{sec:repartitioning}
While selecting the optimal SM split is critical for reducing prefill–decode interference, the timing and stability of repartitioning are equally important. Green Contexts provide logical SM isolation, but switching partition states is not free: transitions can introduce temporary underutilization, overcommitment, or execution stalls.

\vspace{1mm}
\noindent\textbf{Pitfall 1: Synchronous switching.}
A natural design is to synchronize partition changes at global checkpoints. However, as shown in Figure~\ref{fig:switch_sync_design}, this introduces idle “bubbles,” where some streams stall waiting for others to reach the switch point. It hurts SM utilization and inflates latency.

\vspace{1mm}
\noindent\textbf{Pitfall 2: Naive asynchronous switching.}
Letting streams switch independently avoids global stalls, but creates new problems: SM oversubscription or underutilization depending on timing (Figure~\ref{fig:switch_async_design}). This makes system sensitive to transient workload shifts, causing back-and-forth toggling and instability.

\vspace{1mm}
\noindent\textbf{Our design: Buffered asynchronous switching.}
To mitigate these issues, \name{} adopts a buffered asynchronous switching policy. The runtime controller tracks the last-applied SM ratio and only triggers repartitioning when the new target differs by more than a threshold $\delta$. This hysteresis-style buffer (Figure~\ref{fig:switch_async_buffered_design}) smooths out transient fluctuations and suppresses excessive reconfiguration. The logic is embedded in Algorithm~\ref{algorithm:greedy_search}, line 9-13.

This simple mechanism retains the adaptability of fine-grained decisions while avoiding the instability of overreactive switching. Alternative approaches like reducing update frequency or increasing step size proved brittle in practice: the former sacrifices responsiveness; the latter risks overshooting optimal ratios. Buffered switching provides a robust, low-overhead tradeoff.

\begin{algorithm}[t]
    \small
    \caption{\textsc{\name}'s Shortest Prompt First (SPF) Scheduler}
    \label{algorithm:prefill_sch_algorithm}
    
    \begin{algorithmic}[1]
    \State \textbf{Input:} Request queue $Q$, batch limit $B$, age decay factor $\alpha$
    \State \textbf{Output:} The next prefill batch to run
        
        \Procedure{SPF\_Schedule}{\textbf{Queue} $Q$, \textbf{Int} $B$, \textbf{Float} $\alpha$}
            \ForAll{$r \in Q$}
                \State $r.\text{remaining} \gets r.\text{prompt\_len} - r.\text{prefilled\_len}$
                \State $r.\text{score} \gets r.\text{remaining} - \gamma \cdot r.\text{age}$ \Comment{Antistarvation}
            \EndFor
            \State $Q_{\text{sorted}} \gets \Call{SortBy}{Q, r \mapsto r.\text{score}}$
            \State $batch \gets [\ ]$
            \State $total \gets 0$
            \For{$r \in Q_{\text{sorted}}$}
                \If{$total + r.\text{remaining} \leq B$}
                    \State $\Call{Append}{batch, r}$
                    \State $total \gets total + r.\text{remaining}$
                \Else
                    \State \textbf{break}
                \EndIf
            \EndFor
            \State \Return batch
        \EndProcedure
    \end{algorithmic}
\end{algorithm}
\subsection{Phase-Specific Scheduler}
\label{sec:phase_scheduler}
Since \text{\name} separates prefill and decode into concurrent batches, we can further exploit this with phase-specific optimizations. Particularily, we employee tailored scheduler policies that address the distinct latency and resource profiles.

\subsubsection{Prefill Scheduler}
\label{subsec:prefill-scheduling}

TTFT is governed by the prefill stage, which must complete a full forward pass before emitting output. When scheduling prompt requests of varying lengths, naïve policies can introduce head-of-line (HoL) blocking: short prompts are delayed by long ones. This effect significantly impacts latency-sensitive workloads and motivates length-aware scheduling.

At each scheduling tick, the system selects a subset of pending requests whose combined prompt lengths fit within a token budget. Decoupling prefill from decode enables phase-specific scheduling, allowing us to optimize directly for TTFT.

The prompt of each request is known. Hence, we introduce a greedy Shortest Prompt First (SPF) heuristic that ranks requests by an age-adjusted score:
\begin{equation}
\text{score}(r_i) = l_i - \gamma \cdot (t - a_i),
\end{equation}
where $l_i$ is the prompt length, $a_i$ is arrival time, $t$ is the current time, and $\gamma$ controls the anti-starvation trade-off between responsiveness (low $\gamma$) and fairness (high $\gamma$).

Requests are sorted by score and added greedily until the cumulative token limit is reached. This favors short prompts while gradually promoting delayed long requests, achieving a practical balance between latency and fairness. The full procedure is shown in Algorithm~\ref{algorithm:prefill_sch_algorithm}.

While SPF is simple in design, its deployment is facilitated by our system's phase isolation: in monolithic schedulers, prefill prioritization is harder to isolate and tune due to shared queues and coupled resource contention. In \S\ref{subsec:abolation}, we show that SPF significantly reduces TTFT compared to FCFS and helps offset performance regressions when SMs are reallocated to decode under memory pressure.

\subsubsection{Decode Scheduler}
\label{subsec:decode-scheduling}
The decode phase controls TBT. Unlike prefill, decode scheduling operates at a finer granularity, with each request contributing only a single token to the active batch. Thus, we adopt a simple First-Come-First-Serve (FCFS) policy. FCFS ensures fairness, incurs minimal overhead, and avoids token-level starvation. While more sophisticated schemes that considers context window size or memory bandwidth are theoretically possible, we find that their impact is limited in practice as they are considered in SM partitioning. 

\section{Implementation}
We implement \text{\name} on top of vLLM v1-0.8.1~\cite{vLLm-v1}, modifying approximately 6K lines of Python and CUDA/C++ code. Our changes enable phase-separated execution within a single engine or a single GPU, with minimal disruption to the original engine architecture. The maximum batch size and chunk size for prefill of \text{\name} are same as those of vLLM..

\vspace{1mm}
\noindent \textbf{Concurrent Execution.}
We launch prefill and decode phases as separate coroutines, each managing its own GPU stream and scheduler. The main loop coordinates their execution and handles shared metadata updates. Because both phases share worker threads, we guard critical state to prevent inconsistencies during overlapping execution. 

\vspace{1mm}
\noindent \textbf{Per-Phase Scheduler.}
We extend vLLM’s unified scheduler with pluggable logic for phase-specific algorithms. Prefill and decode queues are maintained independently. Both schedulers are configured to use the vLLM's default config. The SPT implementation has the default $\gamma$ set to 15. 

\vspace{1mm}
\noindent \textbf{SM Partitioning and Runtime Switching.} We use  CUDA Green Context~\cite{GreeContext} to partition the SM. Since CUDA Green Context does not provide a Python API, we implement a PyTorch extension using approximately 150 lines of CUDA code to expose this functionality to Python. We leverage this extension to dynamically reassign SM groups at runtime. To avoid reconfiguration overhead, \text{\name} pre-instantiates all partition layouts during initialization and switches among them as decided by the algorithm. The decaying $\lambda$ for each operator in cost model (\S\ref{sec:our_cost_model}) is obtained by profiling prefill and decode offline, and is done for each model and workload configuration. Since SPT scheduler heavily optimizes TTFT, we have a tight 1.1 $\beta$ slack for decode, and 1.3 $\alpha$ slack for prefill. The $KV_{\text{switch}}$ threshold is set to be 70\% of all available KV cache memory.

\begin{table}[t]
\centering
\small
\renewcommand{\arraystretch}{1.2}
\begin{tabular}{l@{\hskip 10pt}l@{\hskip 10pt}rrrrr}
\toprule
\textbf{Dataset} &     & \textbf{Mean} & \textbf{P50}  & \textbf{P95} & \textbf{P99} \\
\midrule
\multirow{2}{*}{\textbf{Long Data Collections}} 
& In  & 5905  & 5461  & 9292 & 9817 \\
& Out & 180  & 159    & 339  & 454 \\
\midrule
\multirow{2}{*}{\textbf{ArXiv Summarization}} 
& In  & 3832 & 3575 & 6460  & 6894 \\
& Out & 200  & 181   & 357  & 443 \\
\midrule
\multirow{2}{*}{\textbf{ShareGPT}} 
& In  & 496  & 432   & 970  & 1367 \\
& Out & 97  & 37  & 383   & 474\\
\bottomrule
\end{tabular}
\caption{\small \textbf{Characteristics of Workloads.} Distributions of input and output lengths of various datasets from different serving scenarios.}
\label{tab:token-stats}
\vspace{-4mm}
\end{table}

\section{Evaluation}\label{sec:evaluation_section}
In this section, we first examine the end-to-end performance of \text{\name} under various workloads. Then, we evaluate the design choices of \text{\name} and show the effectiveness of each component.

\subsection{Experimental Setup}
\label{subsec:experimental_setup}

\textbf{Testbed.} Our evaluations are run on a workstation with Intel Xeon Platinum 8457C CPU (45 cores), two NVIDIA-L20 GPUs with 48GB DDR6 RAM each, and 200GB of CPU memory. The GPUs use driver 570.124.04 together with CUDA 12.8. All benchmarks were executed under PyTorch-2.6.0.

\vspace{1mm}
\noindent\textbf{Model.}
We use Qwen2.5–3B and LLaMA3.1–8B for single-GPU experiments, and Qwen2.5–14B for dual-GPU setups. These popular open-source LLMs span a range of KV cache sizes and compute intensities, enabling evaluation under diverse resource pressures.

\vspace{1mm}
\noindent\textbf{Workloads.}
We construct three workloads to emulate real-world LLM serving, combining datasets with diverse usage patterns and token length characteristics (Table~\ref{tab:token-stats}). Similar to prior work~\cite{vLLM-SOSP23, DistServe}, the arrival pattern of requests is generated by a Poisson process.

\vspace{-1mm}
\begin{itemize}[itemsep=0pt, parsep=0pt, topsep=0pt, partopsep=0pt, leftmargin=*]
\item \textbf{Long Data Collections~\cite{Long-Data-Collections}}: It mixes multi-turn QA and summarization, characterized by long prefill lengths and moderate decode demands. Evaluated on Qwen2.5–3B.
\item \textbf{Arxiv~\cite{arxiv-summarization}}: It uses ArXiv Summarization~\cite{arxiv-summarization} (full paper and abstract pairs) to model long-input, short-output tasks with stable token patterns. Evaluated on Qwen2.5–3B.
\item \textbf{Mixed}: It Combines 60\% ShareGPT~\cite{ShareGPT} (short, interactive prompts) and 40\% Long Data Collections to induce token length and KV cache variability, stressing scheduling and memory. Evaluated on LLaMA3.1–8B and Qwen2.5–14B.
\end{itemize}

\vspace{1mm}
\noindent\textbf{Metrics.}
We report the mean and 95th percentile of three latency metrics: TTFT, TBT, and Normalized Latency.
Normalized Latency is defined as the end-to-end latency divided by the number of output tokens, reflecting per-token serving efficiency across variable-length requests. A well-optimized serving system should maintain low normalized latency under high request loads.

\noindent\textbf{Baselines.}
We compare \text{\name} against four representative LLM serving engines, all configured with the same tensor parallelism, chunked prefill, and scheduling budget for fair comparison.

\begin{itemize}[itemsep=3pt, parsep=0pt, topsep=2pt, partopsep=0pt, leftmargin=*]

\item \textbf{vLLM (\texttt{v1.0.8.1}).}
A throughput-optimized serving engine with FCFS scheduling, continuous batching~\cite{Orca-OSDI22}, Page Attention, and chunked prefill~\cite{ChunkPrefill}.

\item \textbf{FastServe (\texttt{v0.0.8.1}).}
Implements a multi-level feedback queue with skip-join to resolve head-of-line blocking~\cite{FastServe}. We reimplement it atop vLLM due to lack of public code and enable CPU swap (120GB) for each device with recomputation fallback.

\item \textbf{SGLang (\texttt{v0.4.4.post1}) \footnote{For fairness, we select the same evaluation timepoint as vLLM for SGLang; the corresponding commit is 3c09548.}.}
A latency-optimized engine using Radix Attention for KV reuse~\cite{SGlang}. Supports chunked prefill, Page Attention, and FCFS scheduling.

\item \textbf{vLLM-P/D (\texttt{v1.0.8.5}).}
An extension of vLLM with prefill–decode disaggregation via LMCache~\cite{lmcache1, lmcache2, lmcache3}. We evaluate one prefill and one decode instance on separate GPUs to model single-layer PD setups.

\end{itemize}

\begin{figure*}[!t]
\centering
\includegraphics[width=\textwidth]{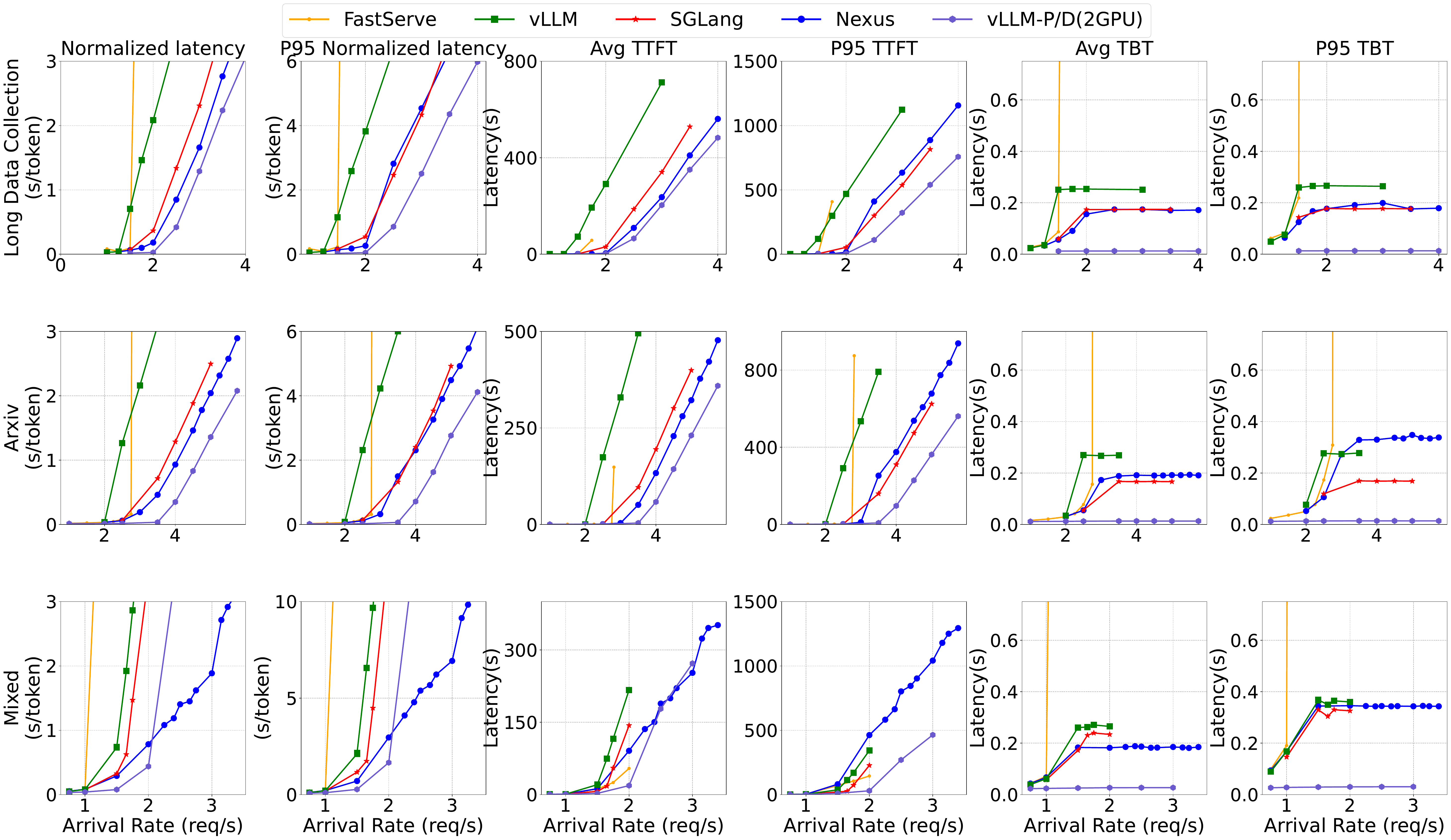}
\vspace{-2mm}
\caption{\small\textbf{End-to-end results on Single GPU.} 
All systems use a single L20 GPU, except vLLM-P/D which uses two. This figure compares three workloads: Long Data Collection and Arxiv use Qwen-2.5–3B(first two rows), and Mixed uses Llama-3–1.8B(third row).  
The first and second columns report the average and 95th-percentile normalized latency—lower is better for throughput. The third and fourth columns report the average and 95th-percentile TTFT, while the fifth and sixth columns show the average and 95th-percentile TBT.}

\vspace{-4mm}
\label{fig:single_gpu_e2e}
\end{figure*}

  


\subsection{End-to-end Performance}
We first evaluate end-to-end performance on a single GPU using three workloads with Qwen2.5-3B and Llama3-1-8B (Section~\ref{subsubsec:single-gpu}). All systems use one L20 GPU, except Dist-vLLM, which uses two. Then, we report multi-GPU results (Section~\ref{subsubsec:multi-gpu}).

\subsubsection{End-to-End Single-GPU Performance}
\label{subsubsec:single-gpu}
Figure~\ref{fig:single_gpu_e2e} reports the end-to-end single GPU performance of \text{\name} and all baselines across three workloads (top to bottom). 
 Figure~\ref{fig:single_gpu_e2e} yields three key conclusions, which we discuss in detail below.

\vspace{1mm}
\noindent\textbf{TTFT.}
As shown in Figure~\ref{fig:single_gpu_e2e} (columns 3–4), \name{} achieves the lowest or near-lowest average TTFT across all workloads. It improves TTFT by 2–20$\times$ over vLLM and up to 1.6$\times$ over SGLang through SPF scheduling and dynamic SM reallocating. The latter reallocates GPU resources at runtime to reduce prefill–decode contention, further amplifying SPF’s benefits. vLLM and SGLang suffer from head-of-line blocking under FCFS, though SGLang fares better due to its optimized runtime. FastServe reduces average TTFT via skip-join MLFQ, but hurts P95 due to deprioritizing long prompts. Compared to vLLM-P/D, which avoids contention by separating phases across GPUs, \name{} matches its TTFT in Mixed Workload and remains within 10\% on Long Data Collections and Arxiv while using a single GPU.

To ensure fairness, \name{} includes a tunable anti-starvation mechanism within its prefill scheduler. Under current settings, it improves P95 TTFT by 2–3$\times$ over vanilla vLLM and narrows the gap with SGLang and vLLM-P/D in Long Data Collections and Arxiv Workloads, while maintaining consistent advantages in average latency. In Mixed Workload, \name{} shows worse tail TTFT due to high prompt length diversity, which increases batching variability and makes it harder to protect long requests without hurting throughput.

\vspace{1mm}
\noindent\textbf{TBT.}
TBT reflects the responsiveness of steady-state decoding, and is particularly sensitive to memory bandwidth and scheduling efficiency. As we can see from Figure~\ref{fig:single_gpu_e2e} (columns 5–6), vLLM-P/D achieves the best average and P95 TBT by fully separating prefill and decode onto dedicated GPUs. Among single-GPU systems, \name{} consistently ranks at or near the top across all workloads.

FastServe degrades sharply under load as it needs to fall back on recomputation. vLLM, which co-schedules prefill and decode, suffers from intra-batch interference, trailing \name{} by 1.24$\times$–1.48$\times$. SGLang improves over vLLM via Radix Attention and runtime optimizations, closely matching \name{} in Long Data Collections  Workload, slightly surpassing in Arxiv Workload, but falling behind in Mixed Workload where prompt diversity intensifies decode imbalance.

While \name{} lags in P95 TBT on Arxiv Workload, this anomaly stems from the workload’s relatively uniform input lengths and moderate output size (Table~\ref{tab:token-stats}), which result in minimal memory pressure. As a consequence, the system’s dynamic SM partitioning is less likely to trigger aggressive decode resource shifts for few tailed requests.

\vspace{1mm}
\noindent\textbf{Throughput.}
To summarize end-to-end performance, we also measure maximum sustainable throughput as the highest arrival rate that each system can handle without violating token latency constraints.

From Figure~\ref{fig:single_gpu_e2e} (columns 1–2), \name{} consistently delivers the highest throughput among single-GPU systems. In Long Data Collections and Arxiv Workloads, it achieves 1.5–1.8$\times$ higher throughput than vLLM and 1.18-1.27$\times$ higher than SGLang, reflecting more efficient resource scheduling under uniform or moderately variable request patterns.

In Mixed Workload, where prompt diversity and scheduling imbalance are most pronounced, \name{} demonstrates its largest gain: it achieves 1.9$\times$ higher throughput than vLLM, 1.8$\times$ over SGLang, and even 1.4$\times$ over vLLM-P/D, despite the latter using two GPUs and full-phase disaggregation. This is driven by \name{}’s head-light design which aggressively prioritizing short requests and adapting GPU resource allocation dynamically, enabling it to avoid contention and return early outputs with minimal delay.

By serving more requests under the same compute budget, \name{} not only improves user-perceived latency but also scales to higher load than other systems, evidencing its strength in practical high-throughput serving scenarios.
\begin{figure}[t]
\centering
\includegraphics[width=\linewidth]{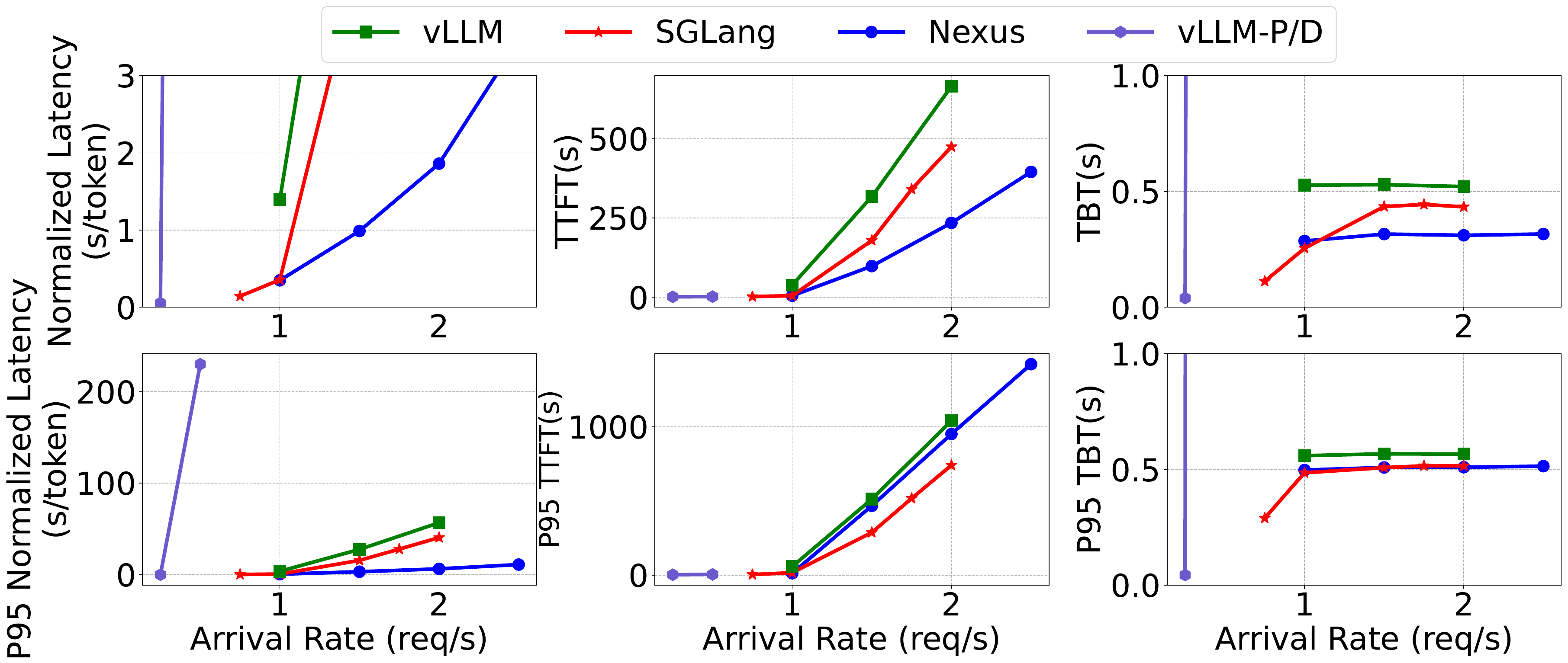}
\vspace{-2mm}
\caption{\small\textbf{End-to-end results for Multi-GPU.} Run using Mixed Workload on two NVIDIA L20 GPUs with Qwen2.5–14B. All systems use two L20 GPU. The top row presents average normalized latency, TTFT, and TBT, while the bottom row shows corresponding P95 metrics.}
\vspace{-4mm}
\label{fig:two_gpu_e2e}
\end{figure}

\subsubsection{End-to-End Multi-GPU Performance}
\label{subsubsec:multi-gpu}Due to space limits, we present multi-GPU results for Qwen-2.5-14B on the Mixed Workload only, as trends on other ones are similar. Since FastServe already performs poorly in single-GPU tests, we exclude it here and compare against stronger baselines: vLLM, SGLang, and vLLM-P/D.


As shown in FIgure~\ref{fig:two_gpu_e2e}, \name{} achieves the highest throughput, 2.2$\times$ over vLLM and 2$\times$ over SGLang, while using the same hardware. These gains come from efficient intra-engine phase separation and the ability to maintain high concurrency without overwhelming shared compute or memory resources.

More importantly, this throughput does not come at the cost of latency. \name{} delivers 2–3$\times$ lower average TTFT and 1.5–2$\times$ lower TBT than vLLM and SGLang, showing strong responsiveness across both prefill and decode phases. At the tail, P95 TTFT is slightly higher than SGLang but matches vLLM, while P95 TBT is nearly identical across all systems.

One surprising result is the poor performance of vLLM-P/D despite its disaggregated architecture. Its aggressive prefill overwhelms the decode stage and saturates the transfer buffer, leading to frequent cache evictions and recomputation. \name{}, by contrast, avoids these issues through adaptive SM partitioning which dynamically changes load to sustain decoding throughput even under pressure.

In sum, \name{} offers the best latency–throughput tradeoff among all baselines, scaling to larger models while preserving per-token responsiveness and avoiding the coordination pitfalls of more fragmented systems.

\begin{figure}[t]
\centering
\begin{subfigure}[b]{0.49\columnwidth}
    \includegraphics[width=\linewidth]{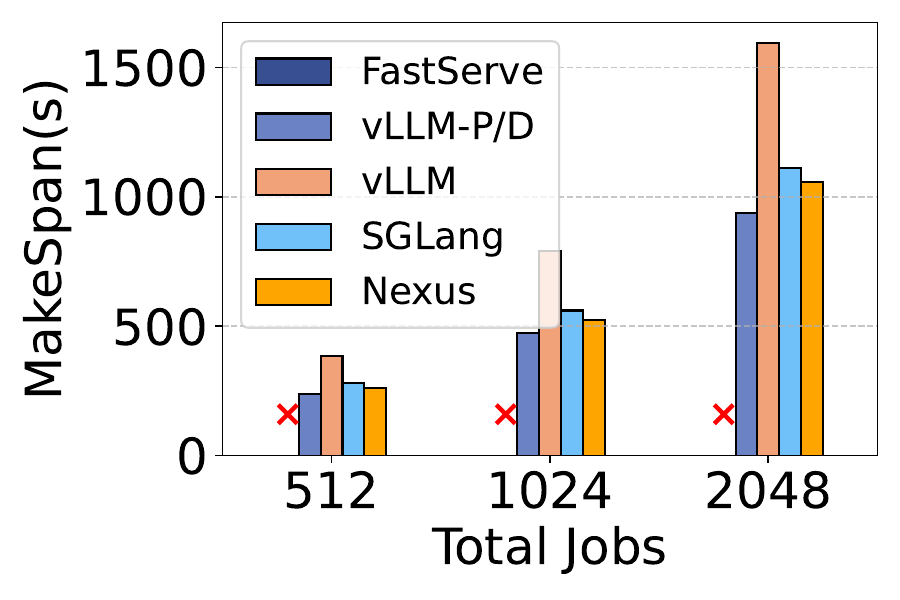}
    \caption{Long Data Collections.}
    \label{fig:workload-i-offline}
\end{subfigure}
\hfill
\begin{subfigure}[b]{0.49\columnwidth}
    \includegraphics[width=\linewidth]{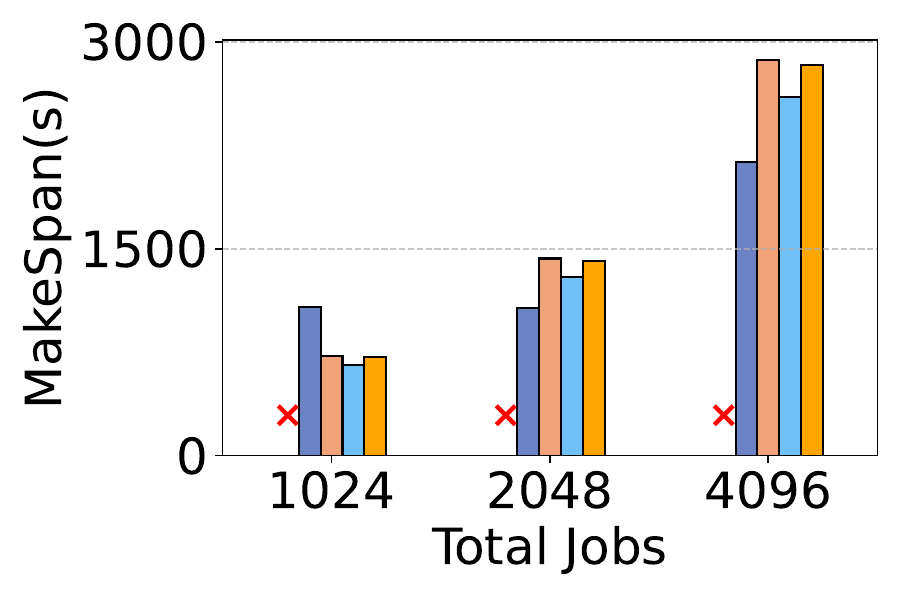}
    \caption{Mixed.}
    \label{fig:workload-iii-offline}
\end{subfigure}

\caption{ \small \textbf{Offline Inference.} Run on a single
L20 under Long Data Collections and Mixed Workloads with
3B and 8B models repsectively. X means timeout. All systems use a single L20 GPU, except vLLM-P/D with two.}
\label{fig:offline}  
\vspace{-4mm}
\end{figure}
\subsection{Offline Inference under Heterogeneous Prompts}
\label{subsec:offline}
In offline settings where requests are handled in large batches, throughput should be prioritized over latency. To evaluate this scenario, we submit all requests at once and measure end-to-end makespan. As shown in Figure~\ref{fig:offline}, \name{} achieves 5–50\% lower makespan than vLLM and SGLang on Long Data Collections, which features uniformly long requests that benefit from phase separation and adaptive GPU resource use. In Mixed Workload, with highly variable input lengths, \name{} still outperforms vLLM by 5\% but lags SGLang by 8–15\% due to its stronger tail control. FastServe times out under both workloads. vLLM-P/D acheives 15\%-35\% lower makespan than \name{} but needs more GPU.

\begin{figure}[!t]
\centering
\begin{subfigure}[b]{0.49\columnwidth}
    \includegraphics[width=\linewidth]{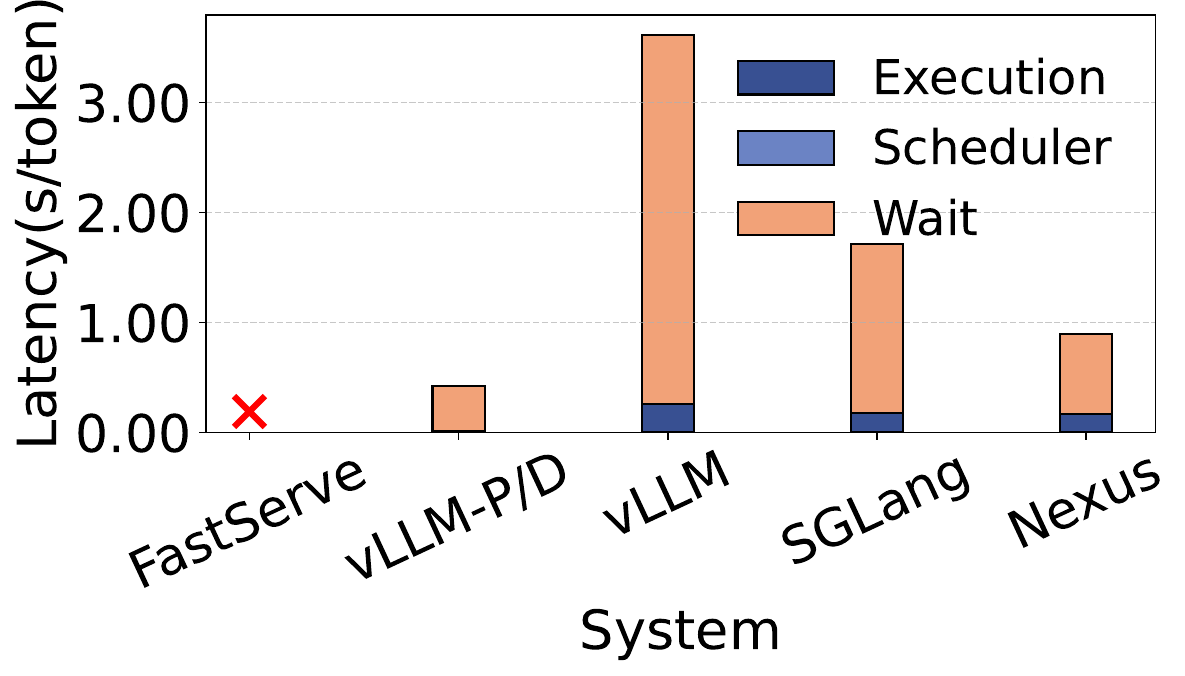}
    \caption{Long Data Collections.}
    \label{fig:workload-i-break-down}
\end{subfigure}
\hfill
\begin{subfigure}[b]{0.49\columnwidth}
    \includegraphics[width=\linewidth]{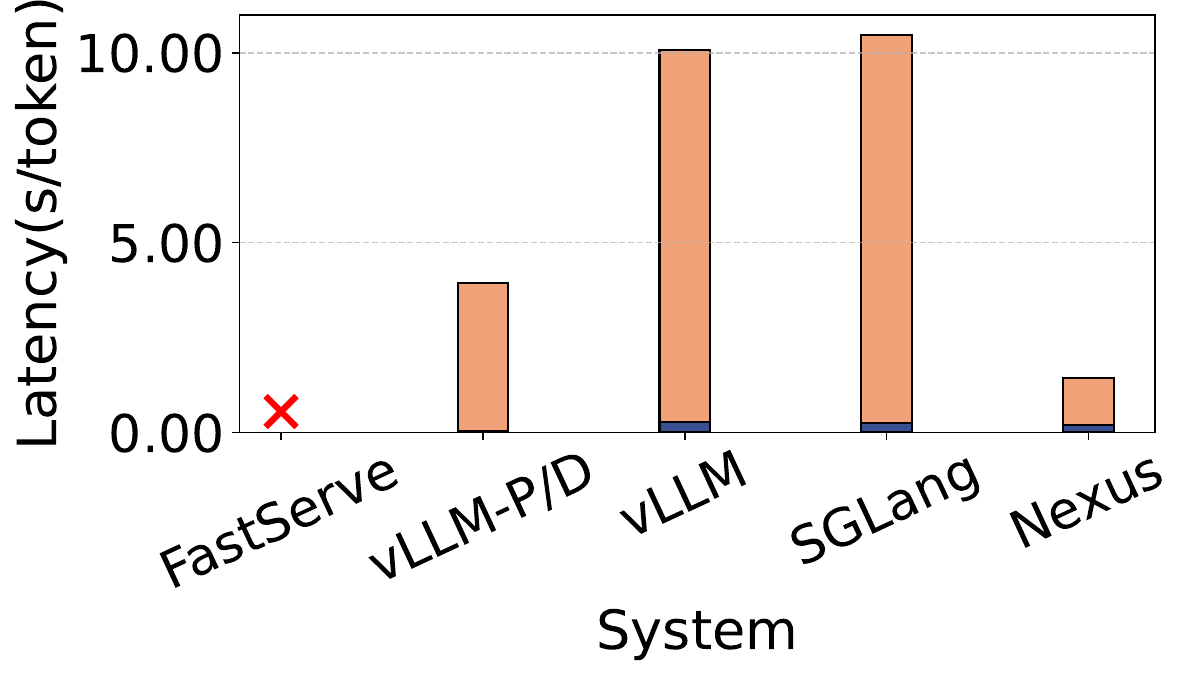}
    \caption{Mixed.}
    \label{fig:workload-iii-break-down}
\end{subfigure}

\caption{ \small \textbf{Breakdown of Inference Overheads.} Run on a single L20 under Long Data Collections and Mixed Workloads with 3B and 8B models repsectively. X means timeout. All systems use a single L20 GPU, except vLLM-P/D with two. }
\label{fig:break_down}  
\vspace{-4mm}
\end{figure}
\subsection{Latency Breakdown}
To better understand the sources of \name{}’s performance gains, Figure~\ref{fig:break_down} decomposes normalized token latency into scheduling, queuing, and execution stages.

\vspace{1mm}
\noindent\textbf{Scheduling Overhead.}
All systems incur minimal scheduling latency. \name{}’s dual-queue design introduces no measurable overhead, confirming its coordination logic is lightweight.

\vspace{1mm}
\noindent\textbf{Execution Time.} Execution latency under \name{} closely matches that of vLLM and SGLang. Although \name{} runs prefill and decode in separate batches, which leads to reading model weights more than once, its batch level separation and dynamic resource use help reduce contention, keeping the overhead low. vLLM-P/D achieves the lowest execution latency due to full disaggregation, but at the cost of using twice GPUs.


\vspace{1mm}
\noindent\textbf{Queuing Delay.}
Waiting time dominates total latency under load, and here \name{} demonstrates its greatest advantage. In Long Data Collections, \name{} reduces waiting time by 4$\times$ over vLLM and 2$\times$ over SGLang. In Mixed Workload, which involves greater request variability, it improves further, achieving 5$\times$ lower wait time than monolithic baselines, and 2$\times$ less than vLLM-P/D. These gains stem from \name{}'s effective shortest-prompt-first scheduling and adaptive GPU resource allocation, allowing it to maintain concurrency without overloading shared resources.

\begin{figure}[!t]
\centering
\includegraphics[width=\linewidth]{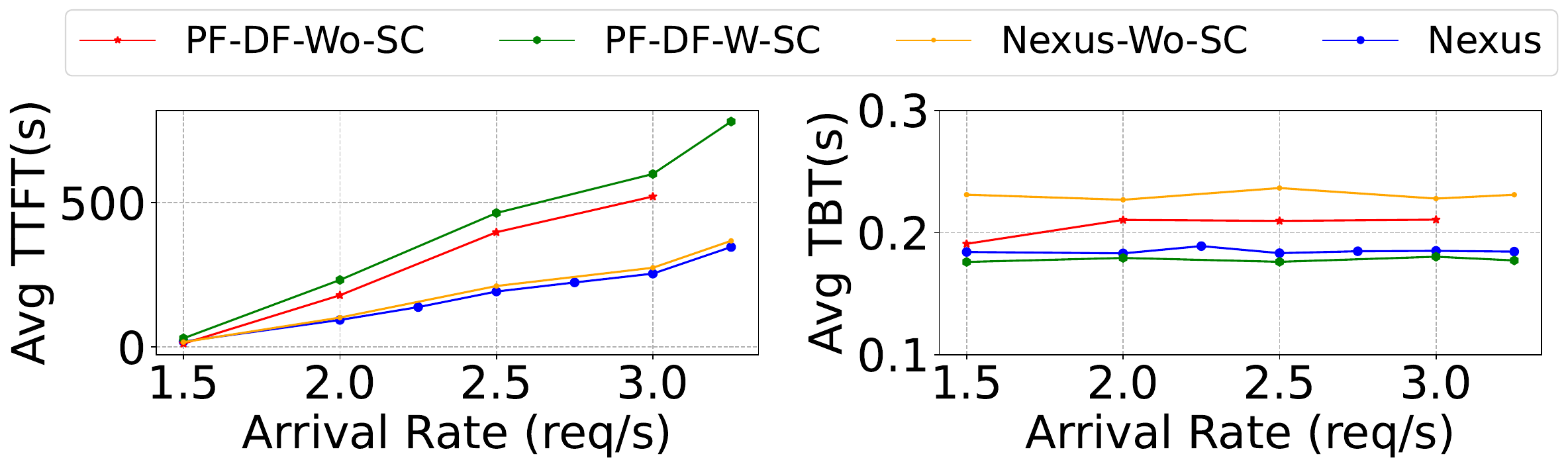}

\caption{ \small \textbf{Ablation Study.} Run with Mixed Workload on Llama3.1-8B using a single L20 GPU.
\textit{PF-DF-Wo-SC} is the intra-engine PD disaggregation that uses FCFS for both prefill and decode scheduling, without dynamical GPU SM changing. 
\textit{PF-DF-W-SC} is the intra-engine PD disaggregation that FCFS scheduling for both prefill and decode but enables dynamical GPU SM changing. 
\textit{\text{\name}-Wo-SC} denotes our system with dynamical GPU SM changing disabled.}

\vspace{-4mm}
\label{fig:ablation}
\end{figure}

\subsection{Ablation Study}
\label{subsec:abolation}
We ablate \name{}’s two core components: (1) dynamic SM changing and (2) phase-specific scheduling, with focus on Shortest-Prompt-First (SPF) for prefill. Figure~\ref{fig:ablation} shows the results.

\vspace{1mm}
\noindent\textbf{Baseline:FCFS Scheduling(Naive Intra-Engine PD Disaggregation).} 
The baseline (PF-DF-Wo-SC) is the intra-engine PD disaggregation that uses FCFS for both prefill and decode without dynamic SM changing. It suffers from head-of-line (HOL) blocking in prefill and persistent resource contention between phases, resulting in poor TTFT and TBT.

\vspace{1mm}
\noindent\textbf{Effect of Dynamical SM Changing.}  
Enabling dynamic SM changing (PF-DF-W-SC) improves TBT by 14\% over the baseline by assigning more compute to decode when GPU memory becomes bottneleck. However, TTFT degrades by 30\% due to delayed prefill execution during decode intervals. This showcases the inevitable TBT and TTFT tradeoff under naive FCFS.

\vspace{1mm}
\noindent\textbf{Effect of Prefill Scheduling (SPF without SM Changing).}  
Applying SPF to prefill (Nexus-Wo-SC) dramatically improves TTFT (up to 90\% reduction over baseline) by mitigating HOL blocking. However, TBT worsens, due to unresolved GPU resource contention between prefill and decode. SPF helps responsiveness, but lacks decode-phase control.

\vspace{1mm}
\noindent\textbf{Combined Design: SPF + Dynamical SM Changing}  
When both techniques are used (\name{}), TTFT improves by 23\% over SPF-only, while TBT drops by 26\%, achieving optimality. Unlike FCFS+switching, TTFT does not regress, as SPF reduces prefill HoL blocking and can benefit from less contention. Dynamical SM changing amplifies gains without introducing new overhead.

\section{Related Work}


\noindent\textbf{Monolithic LLM Serving Systems.}
Early LLM serving engines focus on maximizing throughput and memory efficiency within a unified execution model. Orca~\cite{Orca-OSDI22} introduces continuous batching to reduce head-of-line blocking. vLLM~\cite{vLLM-SOSP23} eliminates KV cache fragmentation via PagedAttention, while SGLang~\cite{SGlang} reduces memory usage through RadixAttention. SarathiServe~\cite{ChunkPrefill} mixes chunks of prefill and decode requests to better utilize GPU resources. However, all of these designs treat prefill and decode as indistinguishable units within a shared queue.
In contrast, \textbf{\name} decouples the two phases at the batching level, enabling independent, phase-specific execution and scheduling while preserving compatibility with existing attention mechanisms.


\vspace{1mm}
\noindent\textbf{LLM Scheduling Frameworks.}
Recent works propose sophisticated schedulers to balance latency and throughput. FastServe~\cite{FastServe} uses MLFQ with skip-join to avoid prompt variance stalls. VTC~\cite{VTC-OSDI'24} and QLM~\cite{QLM-SOCC'24} target fairness and SLO adherence. Llumnix~\cite{Llumnix} accelerates dynamic scaling via migration-aware scheduling. LightLLM~\cite{LightLLm} and Preble~\cite{Preble-ICLR'25} improve memory reuse via future prediction or prompt sharing.
Yet, all treat each request as a monolithic scheduling unit. \textbf{\name} introduces dual schedulers for prefill and decode, each tailored to its phase’s latency sensitivity and compute intensity. This decoupling enables tighter queue control and more efficient GPU resource utilization.

\vspace{1mm}
\noindent\textbf{Engine-Level PD Disaggregation Systems.}
Several systems physically disaggregate prefill and decode across GPUs to better match their resource profiles. Splitwise~\cite{Splitwise} statically assigns phases to different hardware tiers; DistServe~\cite{DistServe} improves TTFT/TBT via tiered GPU scheduling; Mooncake~\cite{MoonCake} serves cached KV blocks from storage to reduce compute load; TetriInfer~\cite{TetriInfer} routes requests by latency class to isolated replicas.
\textbf{\name} achieves similar benefits without incurring cross-engine complexity. By performing lightweight intra-engine disaggregation, it enables low-latency execution and efficient KV reuse within a single serving engine.

\noindent\textbf{Intra-GPU PD Disaggregation Systems.}
There are some prior works on intra-engine PD disaggregation~\cite{drift, bullet}. Drift~\cite{drift} introduces an adaptive gang scheduling mechanism, a contention-free performance model, and an SLO-aware dispatching policy to enable intra-engine prefill–decode separation. Bullet~\cite{bullet} proposes a comprehensive system that includes: (1) a performance estimator for building a profile-augmented analytical model; (2) an SLO-aware scheduler to dynamically balance the compute load between prefill and decode phases; and (3) a resource manager capable of delivering fast yet accurate resource configurations. The key differences between their and \text{\name} are as follows: \text{\name} employs a contention-based cost model to estimate latency, formulates a dual-objective optimization problem, and uses a greedy search to determine the optimal SM partitioning. In addition, our scheduler is phase-aware and explicitly considers the distinct characteristics of prefill and decode stages, which contrasts with others’ SLO-aware scheduling approach.

\vspace{1mm}
\noindent\textbf{Intra-GPU PD Disaggregation Systems.}  
Recent systems like Bullet~\cite{bullet}, Drift~\cite{drift}, and semi-PD~\cite{semi_pd} explore intra-GPU disaggregation to reduce cross-device overhead. Bullet builds profile-augmented latency models and uses SLO-driven feedback loops to tune SM partitioning reactively. Drift applies phase-tagged gang scheduling under a contention-free assumption, combining static modeling with latency-aware dispatch. Semi-PD fits inverse-linear latency curves and adjusts SM ratios through runtime feedback control based on latency violations. \textbf{Nexus} takes a proactive approach. First, it introduces a contention-aware analytical model that explicitly captures both diminishing compute returns and dynamic memory bandwidth interference at the operator level. Second, it formulates intra-GPU resource allocation as a dual-objective optimization problem guided by runtime KV-cache usage and solved via fast greedy search. Third, Nexus uses a one-time profiling pass to calibrate per-kernel latency scaling curves, but avoids offline workload tracing or in-deployment feedback fitting, enabling generalization across dynamic traffic and prompt structures.

\noindent\textbf{GPU Multiplexing.}
Recent work explores fine-grained GPU sharing via spatial or temporal partitioning. NVIDIA MPS~\cite{MPS} and MIG~\cite{MIG} provide coarse-grained isolation, while GreenContext~\cite{GreeContext} enables dynamic intra-process SM partitioning. Systems like GPUlet, Orion, REEF, and Bless~\cite{GPUlet, Orion, REEF, Bless} improve utilization for small models through temporal multiplexing. MuxServe~\cite{MuxServe} proposes spatial-temporal multiplexing to efficiently serve multiple LLMs, while NanoFlow~\cite{NanoFlow} and Liger~\cite{Liger-PPoPP24} introduce kernel level parallelism that enables concurrent execution of compute  bound, network bound, and memory bound operations. PoD~\cite{PoD} fuses prefill attention and decode attention into one kernel to reduce overhead.

\textbf{\name} differs in three key aspects: (1)it disaggregates prefill and decode into different batch and executes these batch concurrently; and (2) it dynamically reallocate SMs across prefill and decode stages in a single engine, adapting to workload shifts; and (3) it introduces intra-engine, phase-specific schedulers for coordinated but decoupled execution. This design enables fine-grained responsiveness while maximizing intra-GPU parallelism.

\section{Conclusion}
\label{sec:conclusion}
While the asymmetric resource demands of prefill and decode phases in LLM inference is well recognized, its implications for intra-GPU coordination have remained underexplored. In this work, we analyze how co-executing these phases under coarse-grained batching strategies leads to resource imbalance and performance interference. To address this, we propose \name{}, a novel SM-partitioned execution framework with dynamic resource allocation and phase-specific schedulers. Our system consistently improves end-to-end throughput and latency across diverse LLM workloads, outperforming state-of-the-art serving systems.

\bibliographystyle{ACM-Reference-Format}
\bibliography{ref}


\end{document}